\documentclass[pdftex,sigconf]{acmart}

\usepackage{amsmath}
\usepackage{url}
\usepackage{algorithmic}
\usepackage{comment}
\usepackage{url}
\usepackage{booktabs}

\usepackage{color}
\usepackage{pbox}
\usepackage{multirow}
\usepackage{xcolor}
\usepackage{subfigure}
\usepackage[figuresright]{rotating}

\DeclareGraphicsExtensions{.pdf,.jpeg,.png}	

\copyrightyear{2020} 
\acmYear{2020} 
\setcopyright{acmcopyright}
\acmConference[CCSW'20]{2020 Cloud Computing Security Workshop}{November 9, 2020}{Virtual Event, USA}
\acmBooktitle{2020 Cloud Computing Security Workshop (CCSW'20), November 9, 2020, Virtual Event, USA}
\acmPrice{15.00}
\acmDOI{10.1145/3411495.3421363}
\acmISBN{978-1-4503-8084-3/20/11}


\definecolor{gatororange}{RGB}{250,70,22}
\newcommand{\lvnote}[1]{\textcolor{purple}{(#1)  --\textit{Luis}}}

\newcommand{\oldtext}[1]{}
\newcommand{\newtext}[1]{{#1}}

\begin{document}

\fancyhead{}
\title{On the Detection of Disinformation Campaign \\Activity with Network Analysis}

\author{Luis Vargas*, Patrick Emami*, Patrick Traynor}
\thanks{* These authors contributed equally.}
\affiliation{
	\institution{University of Florida}
	\city{Gainesville}
	\state{Florida}
}
\email{{lfvargas14,pemami,traynor}@ufl.edu}





\pagestyle{plain}

\begin{abstract}
Online manipulation of information has become more prevalent in recent years as state-sponsored disinformation campaigns seek to influence and polarize political topics \newtext{through massive coordinated efforts. In the process, these efforts leave behind artifacts, which researchers have leveraged to analyze the tactics employed by disinformation campaigns after they are taken down. Coordination network analysis has proven helpful for learning about how disinformation campaigns operate; however, the usefulness of these forensic tools as a detection mechanism is still an open question. In this paper, we explore the use of coordination network analysis to generate features for distinguishing the activity of a disinformation campaign from legitimate Twitter activity. Doing so would provide more evidence to human analysts as they consider takedowns. We create a time series of daily coordination networks for both Twitter disinformation campaigns and legitimate Twitter communities, and train a binary classifier based on statistical features extracted from these networks. Our results show that the classifier can predict future coordinated activity of known disinformation campaigns with high accuracy ($F1 =0.98$). On the more challenging task of out-of-distribution activity classification, the performance drops yet is still promising ($F1= 0.71$), mainly due to an increase in the false positive rate. By doing this analysis, we show that while coordination patterns could be useful for providing evidence of disinformation activity, further investigation is needed to improve upon this method before deployment at scale.}
\end{abstract}
\maketitle

\section{Introduction}
\label{sec:intro}

\begin{figure}
	\vspace{-0.1in}
	\subfigure[]{
		\includegraphics[width=0.9\columnwidth]{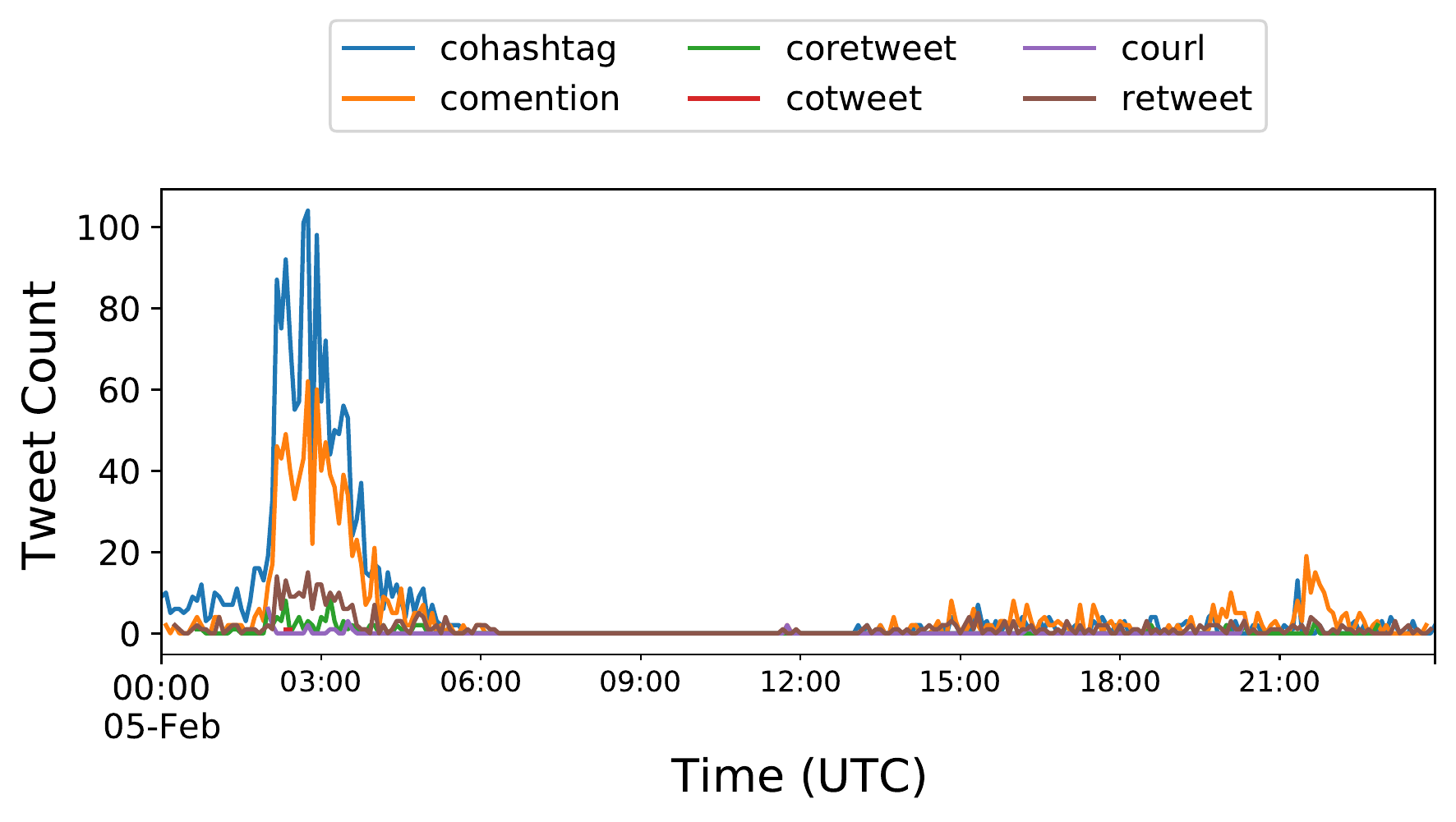}
		\label{fig:sotu-activity}
	}%

	\subfigure[]{
		\includegraphics[width=\columnwidth]{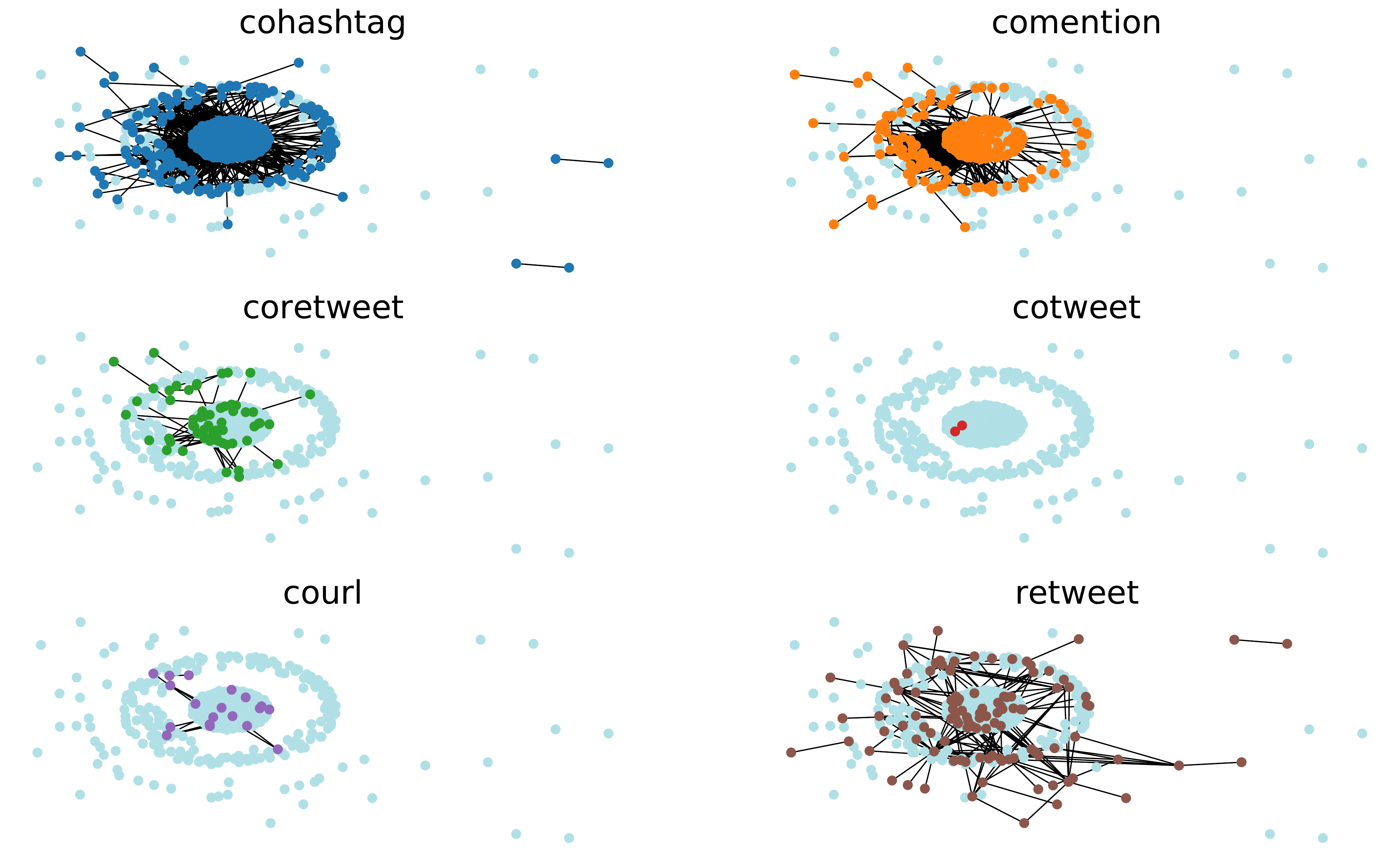}
		\label{fig:sotu-network}
	}
	\caption{\newtext{\textbf{Inauthentic coordination activity detection:} Our work focuses on distinguishing coordinated disinformation activity from other activity found on Twitter. Unlike bot detection, which tries to classify a single node in the networks above, coordinated activity detection requires looking at how accounts interact with each other to push a message in a coordinated manner.}}
	\label{fig:sotu}
\end{figure}

\newtext{

The many forms of social media allow for the rapid and widespread dispersion of information. However, bad actors have exploited the open nature of social media sites to share disinformation with minimal effort. Unlike {\it misinformation}, which refers to the spread of inaccurate news without ill-intent, {\it disinformation} seeks to deliberately spread misleading or inaccurate news for deception and manipulation of a narrative. As Starbird et al. discuss, bad actors have leveraged social media disinformation as a conduit of manipulation to reach and deceive millions of people in the online world as part of their Strategic Information Operations (SIO) campaigns~\cite{starbird2019disinformation}. Many examples of these campaigns have been discovered over the past decade. Some of these can be attributed to state-sponsored manipulation~\cite{bovet2019influence} while others target demonstrations (e.g., \#BlackLivesMatter~\cite{arif2018acting}, \#WhiteHelmets~\cite{wilson2020cross}) and masquerade as grassroots movements (i.e., astroturfing) to appear more genuine to other online users.

In response, many researchers have proposed a range of detection mechanisms for identifying bots, trolls, and botnets (e.g., some examples include~\cite{alizadeh2020content,varol2018feature,varol2017online,yang2019scalable}). While the bot detection has been widely studied for the past decade~\cite{martino2020survey,cresci2020decade,rauchfleisch2020false}, the evolving nature of social media sites has made bot detection less promising due to their focus on identifying automation, lack of providing intent, and only looking at singular accounts. Bot detectors are not applicable for detecting the \emph{inauthentic coordinated behavior} that SIO campaigns show as these use multiple accounts to push their rhetoric at scale. 

We know SIO campaigns exhibit such behavior as many researchers have uncovered this via postmortem analysis of campaigns after takedowns ( see~\cite{keller2020political,zannettou2019let,pavliuc2020,starbird2019disinformation} for examples). To do this, researchers have relied on generating network diagrams to show coordination (e.g., sharing the same message/hashtag/URL) from one account to the next. These networks (along with other signals like context of messages) are then further analyzed to determine the authenticity of SIO campaigns. While coordination networks are an extremely valuable analytical tool, {\it they have yet to be considered as a detection mechanism for modern disinformation.}

Instead of focusing on the detection of accounts, in this paper we address coordinated \emph{activity} classification to provide \emph{evidence} about whether the activity of a group of accounts appears to be part of an SIO campaign (see Figure~\ref{fig:sotu}). To that end, we propose a binary classifier for coordinated activity classification and evaluate whether coordination-network-based features are effective. As such, we consider two relevant prediction tasks in our experiments. First, we analyze a standard time series prediction problem that assesses the ability to classify future coordinated activity given past data from the \emph{same} SIO campaigns. Second, we examine a variant of the first task where classifier must classify activity from \emph{previously unseen campaigns} at test time. Compared to related studies that only use random Twitter activity as baselines for comparison, we argue that comparing coordination patterns of SIO campaigns against other Twitter communities exhibiting varying levels of coordination is more representative of the actual SIO detection problem. Mainly, SIO campaigns and real Twitter communities are composed of users that 1) share similar interests, and 2) are likely to coordinate amongst themselves to push a message to the online world, whether with malicious or legitimate intent. By better understanding the environment in which these coordination-based classifiers will be deployed, we gain insight into how useful they will be for detecting disinformation campaigns. From our large-scale empirical analysis, we observe the following findings:
}
\begin{itemize}

	\oldtext{
    \item {\bf Largest Disinformation Detection Meta-analysis:} We
	measure the effectiveness of previously proposed coordination
	patterns as a way to uncover disinformation campaigns on Twitter by
	examining over \lvnote{UPDATE NUMBERS}51 million tweets from 10 state-attributed campaigns and 
	4 communities with varying levels of benign coordination. 

    \item {\bf Characterize Coordination Patterns of Disinformation
	Campaigns:} Communities with strong ties (especially political
	ones) can coordinate in ways indistinguishable to a network
	analysis-based classifier from disinformation campaigns,
	particularly around major events (e.g., US State of the Union
	address, elections).

}

\item {\bf Coordination Network Analysis as Detection: }
We present the first use of coordination network analysis, commonly used as a forensic tool for postmortem investigation of SIO campaigns, as an activity classification mechanism for modern disinformation campaign data.

\item {\bf Introduce Coordination Activity Classifier: }
We propose a simple binary classifier using coordination network analysis-derived features for SIO-like coordinated activity detection, achieving 0.71 F1 score on out-of-distribution SIO campaigns.
We conduct the largest empirical evaluation of inauthentic coordinated behavior detection to date using over 160M tweets from 10 state-attributed SIO campaigns.

\item {\bf Characterization of Coordination Tactics: }
We analyze the coordination tactics of SIO campaigns and find that such tactics vary within a campaign over time as well as may strongly differ from one campaign to the next. This analysis highlights difficulties faced by coordination activity classifiers, and we suggest future directions to address these concerns.

\end{itemize}


The remainder of the paper is organized as follows:
Section~\ref{sec:relwork} examines related work;
Section~\ref{sec:meth} explains the problem as well as details our implementation and experimental design;
Section~\ref{sec:analysis:results} provides our analysis and findings;
Section~\ref{sec:disc} discusses the broader implications of our work;
and Section~\ref{sec:conc} includes our concluding remarks.

\section{Related Work} 
\label{sec:relwork}

Disinformation campaigns have received much attention from various research disciplines since they play on human influence~\cite{bovet2019influence}, political messaging~\cite{stewart2018examining}, and platform manipulation~\cite{zannettou2017web,wilson2020cross,zannettou2019let}. While online social media platforms like Twitter have presented yet another landscape for the dissemination of disinformation, such platforms have also served as a way to keep a historical record that allows researchers to perform post-mortem analysis of the measures taken by the campaign operators. By focusing on the activity of accounts of one SIO campaign, researchers have been able to infer various campaign characteristics such as: what political leaning the community takes (e.g., left-leaning, right-leaning)~\cite{stewart2018examining}, the divisive content they share~\cite{zannettouChar}, their possible presence in other platforms~\cite{stewart2018examining,zannettou2017web,zannettouChar},  the coordination efforts made by the community~\cite{yang2016social}, or influence~\cite{keller2020political}. More closely related to the work in this paper, other researchers have also looked into cross-campaign comparisons to characterize various disinformation activities~\cite{zannettou2019let,pavliuc2020}. While we use a similar dataset, we focus on detecting SIO coordination activity rather than characterizing campaign content.

Identifying content pollution on social media has become a challenging problem due to the rapid and immense scale at which information diffuses in real-time. As a result, there is an increased interest in automating the moderation of such spaces using carefully designed algorithms~\cite{gorwa2020algorithmic,stringhini2015evilcohort,egele2015towards,echeverri2018lobo}\newtext{---for a recent survey, see~\cite{martino2020survey}}. Such content moderation can be done through bot detection systems like Botometer~\cite{varol2017online,varol2018feature} or disinformation website detection systems like Disinfotron~\cite{hounsel2020supporting}. Additionally, network-based detection mechanisms have also been used to uncover Sybil accounts that wish to abuse online platforms~\cite{zhengsmoke,wang2018graph,cao2014uncovering}. 

Though the bot detection problem involves flagging individual accounts, other detection mechanisms focus on flagging networks based on their behavior. Truthy~\cite{ratkiewicz2011truthy,ratkiewicz2011detecting} is an example of an early automated system for detecting astroturfing campaigns. The primary function of Truthy is to flag networks of users extracted from Twitter as suspicious based on a list of weak indicators. Once flagged, the final determination of whether the coordinated user activity can be considered as astroturfing is left to end-users due to the difficulty of automatically capturing intent.  Though helpful,  the binary classifier with features designed to measure information network connectivity is trained on a small dataset of individual hand-labeled networks obtained via the Truthy web service~\cite{ratkiewicz2011detecting}. Additionally, while Truthy provides evidence that political astroturfing tends to contain patterns of coordination, it lacks rigorous validation on large-scale datasets with non-random legitimate coordination baselines for comparison. Other recent developments in the automatic extraction of Twitter communities likely to be engaged in astroturfing campaigns have mainly focused on tweet coordination networks~\cite{keller2017manipulate,keller2020political,pacheco2020uncovering}, text-based analysis~\cite{keller2017manipulate}, and specific temporal patterns~\cite{pacheco2020uncovering}.

\section{Methodology}
\label{sec:meth}

\newtext{
In this work, we explore the viability of coordination network analysis as a \emph{detection mechanism} rather than as a forensic tool to examine SIO campaigns (e.g.,~\cite{keller2020political,zannettou2019let,pavliuc2020,starbird2019disinformation}). We use labeled data to train a binary classifier to classify a network, which represents the activity of a Twitter community, as SIO-like or not. Instead of using the full histories of activity to create a single coordination network with which to train the classifier as is common for postmortem analysis, we generate coordination networks from daily activity. While immediate detection and suspension of new SIO accounts would be ideal, such a task is non-trivial as there would be minimal-to-no historical evidence of suspicious activity to justify the suspension. Minimizing the amount of time these accounts actively engage with the rest of the world is essential; a key motivation for processing coordination activity over multiple time slices is to be able to classify said activity on a daily basis. For example, this can help a human analyst accumulate evidence to build a case on what may turn out to be an SIO campaign that is progressively becoming more active.

\newtext{
We consider two binary classification variations of increasing difficulty:

\begin{itemize}
	\item[\textbf{\textit{Task 1}}:] \textbf{Time series prediction} Given a training dataset of historical activity from a set of SIO campaigns, at test time we try to classify future activity sampled from \emph{the same} datasets

	\item[\textbf{\textit{Task 2}}:] \textbf{Time series prediction + cross-campaign generalization} Given a training dataset of historical activity from a set of SIO campaigns and Twitter baseline communities, at test time we try to classify future activity sampled from \emph{previously unseen} SIO campaigns and Twitter baseline communities
\end{itemize}

\textit{Task 1} is a standard time series prediction setup similar to one considered in a recent study on detecting social media influence operations~\cite{alizadeh2020content}. We take the daily twitter activity of a subset SIO campaigns and Twitter baseline communities over a span of $N$ days ($t-N,\dots,t-1$) and try to classify new activity on day $t$. This setting evaluates the usefulness of coordination networks extracted from past data at identifying new activity from \emph{known} SIO campaigns. This is less realistic than Task 2, where we do not assume that the test time SIO campaigns have already been detected in the past.  The detection setup of Alizadeh et al.~\cite{alizadeh2020content} broadly differs from our Task 1 in that they try to classify individual user accounts (similar to standard bot detection approaches) using content features as opposed to using coordination activity networks.

\textit{Task 2} is a much more challenging prediction problem and simulates a real-world setting faced by social media sites. Unlike Task 1, at test time we now require the model to classify activity from \emph{previously unseen} SIO campaigns and Twitter baseline communities. This allows us to probe the ability of coordination-network-based detection to generalize across SIO campaigns. We note that Alizadeh et al.~\cite{alizadeh2020content} does not consider this more realistic setting, and we believe this work is the first attempt to analyze the ability of a classifier to extract coordination patterns \emph{across} disparate SIO campaigns.
}}

To summarize, each prediction instance involves collecting coordination activity (e.g., daily activity) from SIO campaigns and Twitter baseline communities over $N$ previous days, training a classifier on this data, and then testing it with activity from a new day and evaluating its ability to properly identify it as SIO or non-SIO.

\subsection{Datasets}
\label{sec:data}







\begin{table}[]
	\centering
	\begin{tabular}{|c|l|c|c|c}
	\hline

	& Campaign (source $\rightarrow$ target) & Accounts & Total Tweets \\
	\toprule

	\parbox[t]{2mm}{\multirow{10}{*}{\rotatebox[origin=c]{90}{Twitter SIO}}}  
    &Unknown $\rightarrow$ Serbia                              & 8.2K& 43.07M\\
	&Turkey $\rightarrow$ Turkey                               & 6.3K& 36.95M\\
	&Saudi Arabia $\rightarrow$ Saudi Arabia                   & 4.5K& 36.52M\\
	&Saudi Arabia $\rightarrow$ Geopolitical                   & 5.9K& 32.06M\\
	&Egypt $\rightarrow$ Iran/Quatar/Turkey                    & 2.2K& 7.96M\\
	&Russia (internal) $\rightarrow$ Russia                    & 1.0k& 3.43M\\
	&Indonesia $\rightarrow$ Indonesia                         &  716& 2.7M\\
	&Honduras $\rightarrow$ Honduras                           & 3.0k& 1.1M\\
	&China (2020) $\rightarrow$ Hong Kong                      &23.75k& 349k\\
	&Russia (external) $\rightarrow$ Russia                    &  60& 40k\\
	\midrule
\parbox[t]{2mm}{\multirow{4}{*}{\rotatebox[origin=c]{90}{Baselines}}}  &	UK Parliament (political)              & 568& 1.52M\\
&	US Congress (political)                          & 527& 1.47M\\
&	Academics (non-political)                     & 818& 1.34M\\
&	Random (non-political)                        & 543& 1.21M\\
\bottomrule

	\end{tabular}
    \caption{High level statistics of the Twitter disinformation campaigns and manually collected political and non-political communities used in our classifier analysis. Note that while SIO communities have more accounts than the baseline, coordination amongst all of those accounts does not occur simultaneously (See Figure~\ref{fig:activity}).} 
 	\label{tbl:datastats}
\end{table}

In this section, we discuss how we collected the ground truth Twitter activity for SIO campaigns as well as four Twitter baseline communities. We use the coordination networks extracted from these communities as a way to understand if coordination patterns are unique to SIO campaigns. We acknowledge that SIO campaigns also exist in other online platforms (e.g., Facebook, Reddit); We chose to base our analysis on Twitter since many detection tools have already been proposed for this platform and Twitter also provides SIO-related data to the open public for research purposes. 

\subsubsection{Twitter SIO Campaigns}

For SIO ground truth, we accessed the data archives of Twitter's Information Operations Election Integrity reports~\cite{twitterElectionHub}. \newtext{We only use SIO campaigns that were active between January 1st, 2018 to December 31st 2019 since these campaigns had a temporal overlap with the baselines we collected.} Each data archive contains all tweets, media shared, and descriptions of accounts that were part of a potentially state-backed information operation campaign. For each campaign, we grouped the archives based on the state-sponsor attribution (source) Twitter provided as well as which countries (targeted) were attacked by the SIO campaign\footnote{Exact mapping based on the archive to state-sponsored attribution can be found in Appendix~\ref{appendix:mapping}.} (e.g., China state-sponsored targeting Hong Kong). All coordination graphs and features mentioned in Section~\ref{sec:features} were generated based on this combination of data archives in their respective source/target campaigns. The high-level statistics of the SIO campaigns are shown in Table~\ref{tbl:datastats}.

\subsubsection{Twitter Baseline Communities Activity}
To compare SIO activity to potentially legitimate coordination present on Twitter, we collected activity from four communities that have varying levels of coordination amongst their members by using the Twitter API, which provides an account's most recent 3200 tweets. These four baselines are broadly categorized as either political or non-political Twitter communities. We make this categorization to highlight the similarities that SIO campaigns may have to certain Twitter communities that engage in politically-charged discussions and other communities that mainly discuss non-political topics. \newtext{Additionally, the four baselines are meant to show community behavior. As such, the baselines are not necessarily representative of the average Twitter user.}

For the political baselines, we collected the activity of accounts from the members of the United States Congress as well as the United Kingdom's Members of Parliament (MPs). These two communities were chosen since their members are likely to discuss political topics that have polarizing viewpoints as well as coordinate amongst themselves to pass legislation. 

The two non-political communities we collected were accounts of academic/professional security researchers and a random set of accounts. We chose academics as an example of a legitimate Twitter community since many members are likely to show coordinated behavior (e.g., tweet about a conference, promote a paper) and engage in relatively fewer political discussions. We claim that our academic baseline is representative of other Twitter communities composed of accounts that share a similar interest, such as other academic fields or users that are fans of the same sports team. In addition to the academics baseline, we collected a random set of accounts as previous studies have done. We collect this common baseline by first querying tweets that are part of a randomly chosen trending hashtag, then randomly picking one account from the results, and finally doing a random walk from account to account based on whom they follow. This random baseline is expected to have minimal (if any) shared interests or coordination. Statistics on our set of baseline communities are provided in Table~\ref{tbl:datastats}.

\begin{figure} 
  \includegraphics[width=\columnwidth]{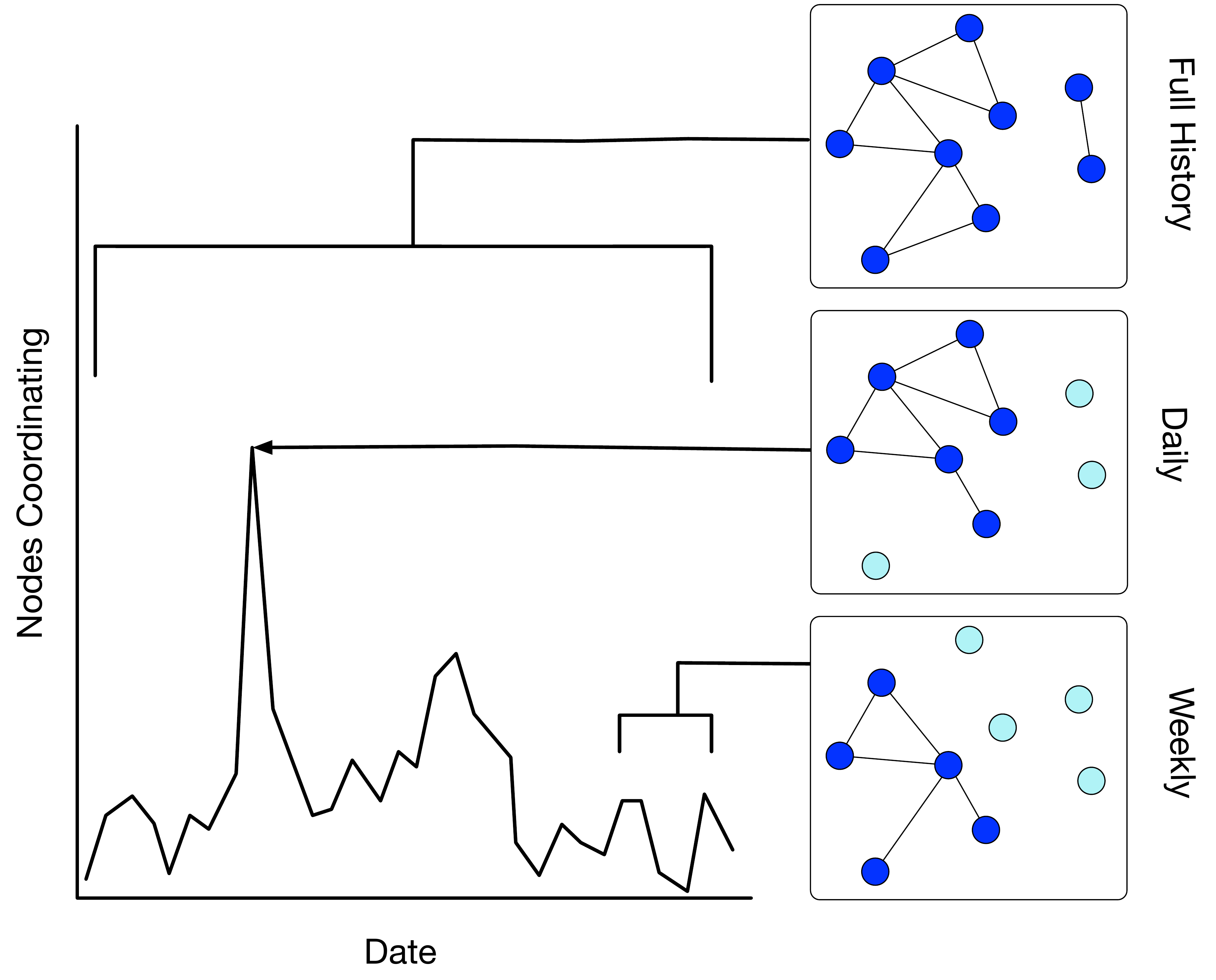}
  \caption{Looking at the coordination patterns of a campaign at specific time periods can reveal ``events'' where the campaign pushes their messages. \newtext{Extracting network statistics of said events can be used to distinguish SIO-like activity and regular coordination found on Twitter.}}
  \label{fig:features} 
\end{figure}
\newtext{Note that the baselines mentioned show varying levels of organic activity. For example, accounts in academic and random baselines are assumed to be controlled by regular people. However, accounts present in both the US Congress and UK Parliament are likely to employ some automation as they may post tweets at pre-specified times and may also have staffers controlling the account. While the behavior may appear to be automated, the activity presented is legitimate as Twitter both verified many of these accounts as well has not suspended them for their activity.}

\newtext{
\subsubsection{Class Imbalance} To help achieve our goal of making our experiments as organic as possible, we would like to emulate that there are many more legitimate communities interacting on Twitter on a given day than there are SIO campaigns. Therefore, for our experiments, we oversample the activity from the four legitimate communities to create a 1:9 class imbalance (2 SIO campaigns to 18 non-SIO communities) as well as create a 1:3 imbalance (5 SIO campaigns to 15 non-SIO). Apart from being more balanced, the latter allows us to investigate training a classifier across a large number of SIO campaigns. 
}

\subsection{Coordination Network Features}
\label{sec:features}

We now discuss how we extract feature vectors from coordination found in SIO campaigns and the baseline communities. We also discuss the implementation details of our classifier.

\newtext{
To obtain the coordination network features, we extract coordination patterns that have been used by other researchers to analyze SIO campaigns~\cite{ratkiewicz2011truthy,ratkiewicz2011detecting,keller2020political,keller2017manipulate}. However, instead of using a single monolithic coordination network over the entire lifespan of a campaign as is common for forensic analysis, we aggregate activity by day and by week (Figure~\ref{fig:features}). Experiments denoted as \emph{daily} involve training classifiers on coordination features derived from each individual day's coordination activity. The \emph{weekly} experiments differ in that the feature vector corresponding to day $t$ contains an aggregate of all coordination from the last seven days $(t-6,\dots,t-1,t)$. In practice, we keep the unique set of accounts present in the week and take the union of all of the edges from each daily coordination network.}

\newtext{
In Figure~\ref{fig:features}, we show the intuition behind using daily and weekly coordination. Compared to a single monolothic network, daily and weekly coordination allows us to frame the problem as time series prediction and also helps distinguish days/weeks where an SIO campaign is actively trying to push a message. Notice that joining all daily coordination networks yields the monolithic coordination networks that previous researchers have used to examine SIO campaigns.
}


Our methodology focuses on the coordination behavior of a community. Coordination of a community can be represented as a network where nodes are accounts and edges represent some prescribed coordination pattern that may exist between two accounts. In our case, we consider six coordination patterns that have been previously proposed as indicators of SIO~\cite{ratkiewicz2011truthy,ratkiewicz2011detecting,keller2020political}:

\begin{itemize}
	\item {\bf retweet:} an account retweeting another account in the same community~\cite{keller2020political,ratkiewicz2011detecting}
	\item {\bf co-tweet:} two accounts tweeting (not retweeting) the same content (url excluded) within a time threshold~\cite{keller2020political}
	\item {\bf co-retweet:} two accounts retweeting accounts that may/may not be part of the community within a time threshold~\cite{keller2020political}
	\item {\bf co-hashtag:} two accounts tweeting the same hashtag within a time threshold~\cite{ratkiewicz2011detecting}
	\item {\bf co-mention:} two accounts mentioning the same user within a time threshold~\cite{ratkiewicz2011detecting}
	\item {\bf co-url:} two accounts sharing the same url within a time threshold~\cite{ratkiewicz2011detecting}
\end{itemize}

While some of the patterns mentioned above are common behaviors in Twitter, Keller et al. ~\cite{keller2020political} argues that accounts tweeting/retweeting messages with similar content within a short time frame can be seen as suspicious activity, and so we use a time threshold for all types except for the retweet networks. Retweeting is a simple way to spread false information and, more generally, to signal-boost a specific message. Time thresholds were not enforced for the retweet coordination pattern in previous studies~\cite{keller2020political,ratkiewicz2011detecting} and hence we follow suit. Conversely, two or more tweets using the same hashtag may not necessarily mean that these accounts are coordinating. However, if those accounts tweet the same hashtag within a time threshold (e.g., 1 minute) of each other, then the behavior starts to appear more suspicious. The timing is important since short time between two messages with the same content could mean automation or an individual controlling multiple accounts. For all coordination patterns, the pattern is defined by what they share and the time between messages (except retweet).

To generate the daily feature vectors that we use to train the classifier, we first collect all the tweets from a community within a given time period. Second, for each coordination pattern mentioned above, we generate the coordination network using the tweet content and time to determine which accounts require an edge between them based on the specified coordination requirements. Once the coordination networks are generated, we extract the seven statistical properties mentioned in Table~\ref{tab:features} for each of the six networks that measure the amount of activity (e.g., number of nodes and edges) and the connectivity (e.g., average connected component sizes, average node degrees). We concatenate all the high-level statistics into a 42-dimensional vector (seven metrics for each of the six coordination networks) to train/test the classifier.

We note that results based on co-url activity are limited due to link shortener/redirection services. 
While for the collected baselines, we can extract the exact links that were posted by the user, the URLs that are part of the Twitter SIO archives do not have such information present. Instead, a non-negligible amount of URLs are shown as their shortened URL version (e.g., \texttt{bit*ly}, \texttt{dlvr*it}). While the correct approach to solve this issue is to look for the redirection of the shorten URLs manually, we noticed that some end domains no longer existed at the time of our analysis. Thus, instead of redirecting us to the originally posted URL, we get redirected to the domain registrar (e.g., \texttt{www*hugedomains*com}). Basing edge creation on these misleading redirects could add non-existent edges to the network. As such, we decided to be conservative with our approach and use the URLs found in the dataset instead of the redirected values.

\begin{table}[t] 
	\begin{center}

		\begin{tabular}{c|l}
			\toprule
			& Coordination Network Features \\
			\midrule
			\texttt{nodes} & \# of nodes \\
			\texttt{edges} & \# of edges \\
			\texttt{largest\_cc} & Size of the largest connected component \\
			\texttt{mean\_cc} & Avg.  size of connected components \\
			\texttt{std\_dev\_cc} & Std. dev. of the sizes of connected components \\
		\texttt{mean\_deg} & Avg. node degree  \\
		\texttt{std\_dev\_deg} & Std. dev. of node degree \\ \bottomrule
		\end{tabular}
		\vspace{0.1in}
		\caption{The seven features extracted from each of the six types of coordination networks from the SIO campaigns and baseline communities. We concatenate them to form 42-dimensional feature vectors. These features were chosen since they appear in previous work~\cite{ratkiewicz2011detecting}.}
		\label{tab:features}
	\end{center}
\end{table}

\subsection{Classifier Implementation}
\label{sec:implementation}

For the binary classifier, we use a Random Forest (RF)~\cite{breiman2001random} with 100 trees implemented using scikit-learn~\cite{scikit-learn}. RFs are widely considered to be a strong baseline classifier and have recently been employed for bot detection~\cite{yang2019scalable}, social influence campaign detection~\cite{alizadeh2020content}, and disinformation website detection~\cite{hounsel2020supporting} at scale. Compared to deep learning approaches, RFs learn similar highly nonlinear decision boundaries but produce more interpretable results, are more robust to unnormalized features, and generally handle low to medium amounts of training data better. \newtext{We do not do a grid search over RF hyperparameters since our primary goal is testing whether an out-of-the-box classifier can distinguish SIO activity from legitimate activity using coordination network features.}

\newtext{
Due to the class imbalance in our datasets, the most appropriate way to evaluate the classifier's performance is by inspecting precision and recall scores.  We use majority voting across the ensemble of 100 trees as a decision boundary to compute F1 measure. Majority voting is the standard decision rule for RFs, and generally trades off precision (P) and recall (R) quite well. For completeness we include receiver operating characteristic (ROC) and P-R curves in the appendix.

Our Task 1 experiments are designed as follows. First, we randomly select two SIO campaigns and oversample the four non-SIO baselines to get 18 non-SIO total---creating a positive-negative label ratio of 1:9. For each day between January 1st, 2018 and December 31st, 2019, we collect the activity over $N$ previous days from the 20 sources. For the \emph{daily} experiments, a RF classifier is trained on the $20 \times N$ coordination activity feature vectors, and then evaluated on the 20 feature vectors from the held-out day. For the \emph{weekly} experiments, we train the model on days $(t-N, \dots, t-1)$ and test on day $(t+6)$ to avoid any leakage between train and test. We enumerate all pairs of SIO campaigns out of the ten ${10 \choose 2}$, resulting in 45 distinct pairings.  We also consider a scenario with a 1:3 label ratio, where we uniformly sample 100 sets of five SIO campaigns out of ten and oversample the four baselines to get 15 total. Note that the coordination activity seen at test time comes from the same campaigns and communities we train on in Task 1.

The main difference between Task 1 and Task 2 is that we now sample the test time activity from an unseen SIO campaign and baseline community. We enumerate all groups of three SIO campaigns out of ten ${10 \choose 3}$, resulting in 120 distinct triplets where the model trains on two campaigns and evaluates on the third. For each of the 120 SIO sets, we randomly select three of the four baselines and oversample them to achieve the same label ratios as before. To maintain the label ratios at test time, we flip a biased coin with probability 1/9 or 1/3; if it lands on heads, we test the classifier on the SIO data, if tails, we test it on the non-SIO data. 

To select $N$, the number of days used in the sliding window for time series prediction, we set aside all data between January 1st, 2018 to March 31st 2018 to use as a tuning set. Out of $N = 1,\dots,60$, $N=60$ attained the best results on Task 1 over the tuning set showing monotone improvements as N increased, and hence we use it for all experiments. For simplicity, we only use a strict coordination time threshold of 1 minute (e.g., we only add an edge between two nodes in the co-tweet network if two accounts both make an identical tweet at most minute apart).}

In total, from the 166.3M tweets across all disinformation campaigns and baseline communities, 157K coordination networks were generated to represent coordination activity. Computing these networks took 3 days on a 40 core/512GB RAM server.

\section{Results}
\label{sec:analysis:results}

\begin{figure*}[th]

	\subfigure[Saudi Arabia  $\rightarrow$  Geopolitics]{
		\includegraphics[width=0.3\textwidth]{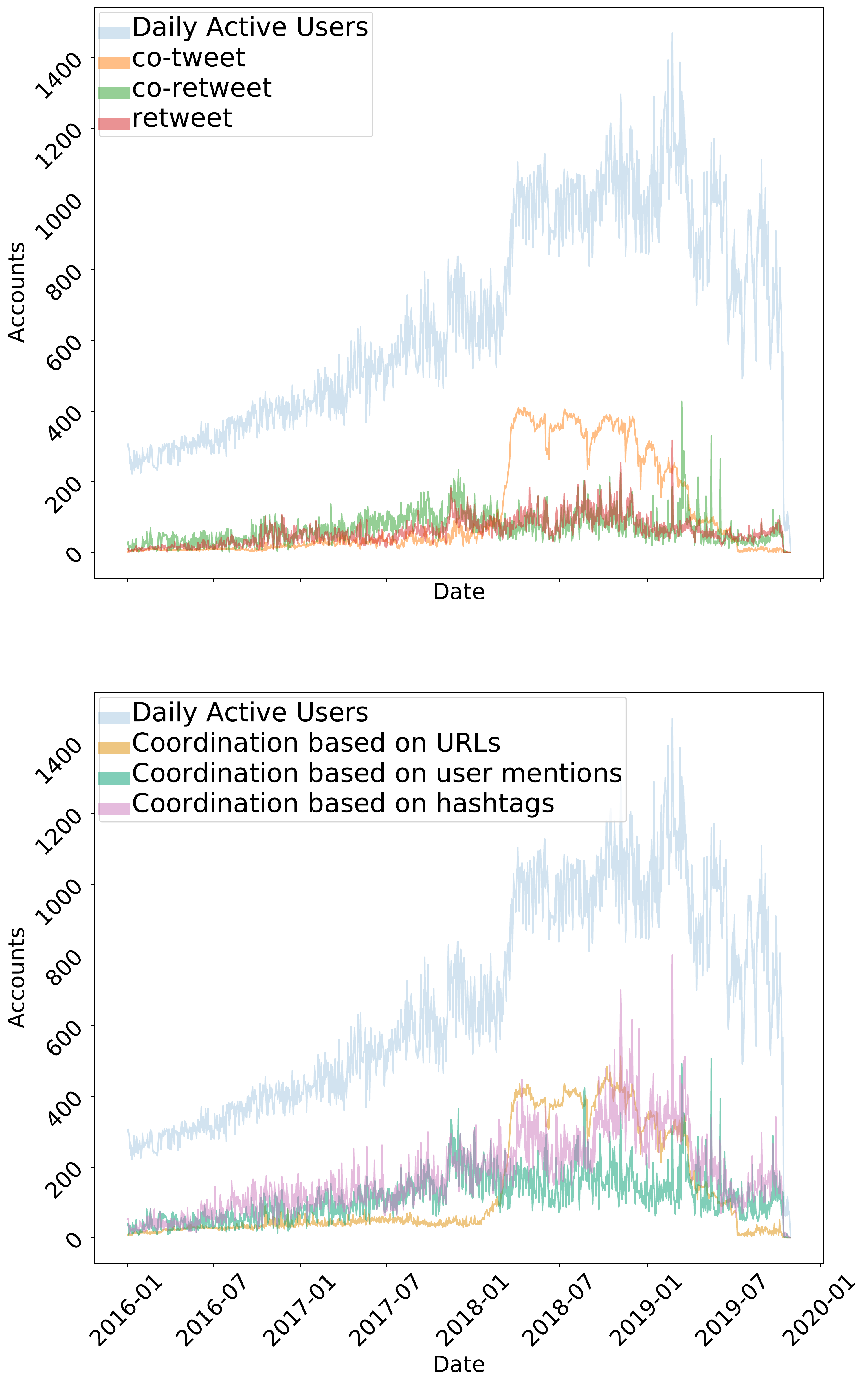}
		\label{fig:saudi}}
	\subfigure[Turkey $\rightarrow$  Turkey]{
		\includegraphics[width=0.3\textwidth]{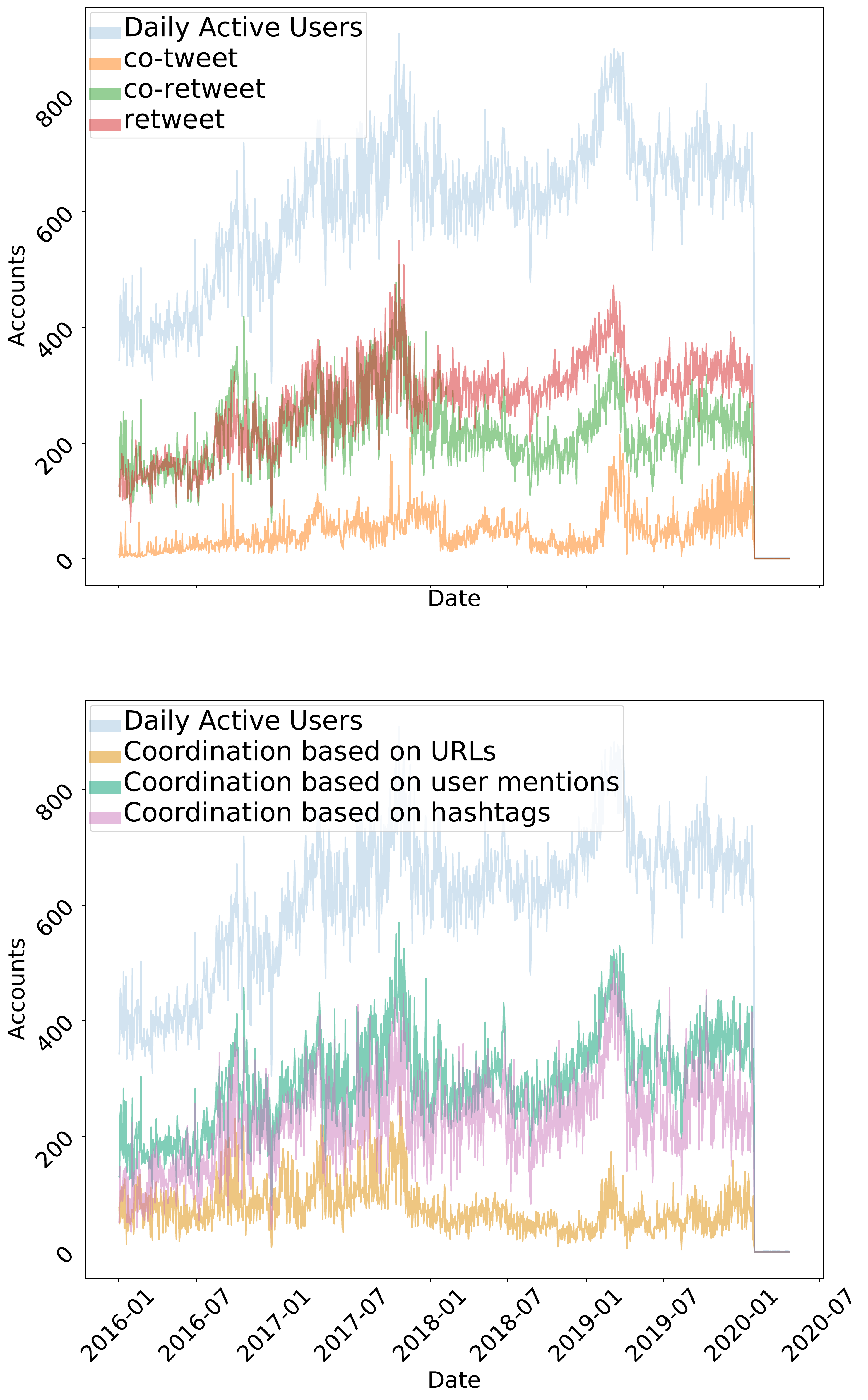}
		\label{fig:turkey}}
	\subfigure[US Congress]{
		\includegraphics[width=0.3\textwidth]{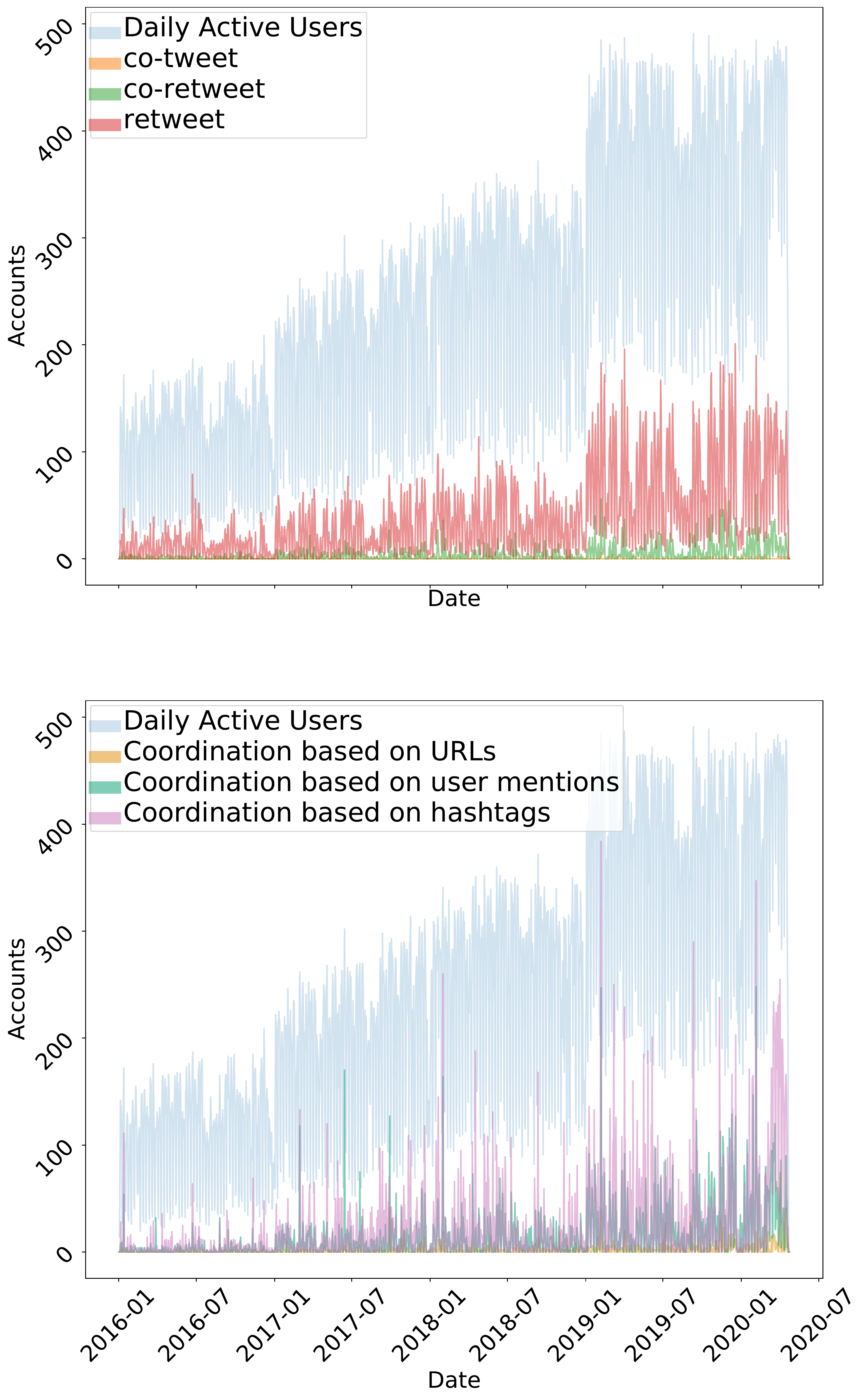}
		\label{fig:congress}}
	\caption{Examples of coordination activity for three communities (two disinformation campaigns/one baseline). \newtext{Each time series represents the number of accounts (nodes in networks) coordinating in a daily basis using a specific coordination pattern.} Campaigns display different behavior characteristics such as intermittent use of co-tweeting (Saudi Arabia) or a diverse but consistent set of coordination patterns (Turkey). \newtext{For the baselines, political communities show more daily coordinated activity than non-political (shown in Appendix~\ref{appendix:mapping})}. This image is best seen in colors.
	}
	\label{fig:activity}
\end{figure*}

\newtext{
We conduct a series of empirical studies to analyze the tasks posed in Section~\ref{sec:meth}.  First, we qualitatively characterize how coordination changes and differs across SIO campaigns and Twitter community baselines. Then, we measure the ability of our binary classifier to correctly identify daily/weekly coordinated activity in Task 1.  For Task 2, we run a similar experiment to \textit{Task 1} but hold out SIO campaigns to simulate discovery of unknown SIO activity. We interpret the trained models by visualizing the coordination feature vectors using t-SNE~\cite{maaten2008visualizing} and discuss feature importance using SHAP plots~\cite{NIPS2017_7062}. Finally, we do a deep dive into misclassifications to attempt to explain why they occurred.
}


\subsection{Characterization of Coordination}
\label{section:analysis:high-level-char}

\subsubsection{Setup}
Before we analyze the tasks described in Section~\ref{sec:meth}, we first do a high-level characterization of coordination patterns in our data. We generated the coordination networks for each pattern discussed in Section~\ref{sec:features} using daily aggregation with a 1-minute time threshold; we show some example patterns in Figure~\ref{fig:activity}. We counted how many accounts participated in each coordination pattern in a given day (i.e., the number of nodes in the coordination network). Note that each column represents the coordination patterns for one campaign, and the scale of the y-axis are different. Due to space limitations, we show the rest of the coordination patterns in Appendix~\ref{appendix:mapping}.

\subsubsection{Key Results}
We find that the usage of each type of coordination pattern differs from one community to the next. From Figure~\ref{fig:activity}, we see that coordination in SIO campaigns appear to be more orchestrated than legitimate activity. For example, in the Saudi Arabia campaign (Figure~\ref{fig:saudi}), they appear to make use of co-tweeting (yellow) for most of 2018 but then stop. They then mainly focus their coordination efforts on simpler tactics like sharing the same URL (orange) or hashtags (purple). Conversely, coordination for the Turkey campaign (Figure~\ref{fig:turkey}) appears to have a steady usage of coordination types across their entire life cycle. These SIO campaigns show relatively constant coordinated activity, whereas other SIO campaigns (Appendix~\ref{appendix:mapping}) show different burst-like behaviors (i.e., intermittent levels of high activity and little-to-no activity). 

For the Twitter community baselines, we can see that retweeting is the most common coordination pattern. This is not surprising as retweeting is one of the simplest actions that a user can take on Twitter as it only needs one click. Additionally, accounts within a community may desire to signal boost each other by retweeting if they share similar goals (e.g., legislative). The amount of coordination also varies for the baselines, with political communities (the UK Parliament and the US Congress) showing more coordination than our non-political communities (Random and Academics) (see Appendix~\ref{appendix:mapping}).

\subsubsection{Takeaways}
As one might expect, coordination appears to be more prevalent and orchestrated in SIO campaigns than in the Twitter community baselines. However, the type of coordination, as well as the behavior based on it, differs for each campaign. \newtext{We hypothesize that} this variability will make it hard \newtext{to learn generalizable patterns for effective detection of previously unseen campaigns. We observe that the amount of coordination tends to increase over time until the campaign is taken down, which a sliding window setup should help address}. 


\subsection{Main experiments}
\label{sec:analysis:classification}

\begin{table*}[]
	\centering
	\begin{tabular}{@{}cccccc@{}}
	\toprule
	Task & Class Ratio & Aggregation & F1  ($\uparrow$)                & Precision ($\uparrow$)          & Recall ($\uparrow$)             \\ \midrule
	1 & 2 : 18   & Daily    & 0.879 (0.846,0.912)           & 0.959 (0.944,0.975)          & 0.853 (0.809,0.897)          \\	
     1 & 5 : 15   & Daily    & 0.897 (0.888,0.906)           & 0.955 (0.949,0.960)          & 0.870 (0.856,0.884)          \\
	1 & 2 : 18   & Weekly   & 0.969 (0.965,0.972)           & 0.980 (0.976,0.984)          & 0.971 (0.966,0.975)          \\
	1 & 5 : 15   & Weekly   & \textbf{0.980 (0.979,0.981)}  & \textbf{0.985 (0.984,0.986)} & \textbf{0.982 (0.981,0.983)} \\
	\midrule
	2 & 2 : 18   & Daily    & 0.516 (0.471,0.561)   & 0.445 (0.366,0.523)          & 0.724 (0.710,0.738)              \\
	2 & 5 : 15   & Daily    & 0.671 (0.654,0.688)   & \textbf{0.641 (0.600,0.682)}  & 0.732 (0.714,0.750)   \\
	2 & 2 : 18   & Weekly   & 0.397 (0.360,0.433)           & 0.283 (0.241,0.324)    & 0.715 (0.694,0.735)   \\
	2 & 5 : 15   & Weekly   & \textbf{0.709 (0.687,0.730)}      & 0.602 (0.562,0.641)   & \textbf{0.877 (0.850,0.903)}      \\ \bottomrule
	\end{tabular}
	\vspace{0.1in}
	\caption{\newtext{Coordinated activity detection results for \textit{Task 1} (time series prediction) and \textit{Task 2} (time series prediction + cross-campaign generalization). We show 95\% CIs in parenthesis. We find that the combination of a historical window of $N=60$ days with weekly aggregation successfully classifies future activity in the short-term if the coordination patterns during test time are identical to those seen during training. However, for the more realistic setting of \textit{Task 2}, classifier performance drops because the behavior of SIO campaigns and legitimate during test time are from novel sources. Best results on this task are achieved with weekly aggregation and a more balanced dataset with a sufficient number of SIO campaigns to extract general patterns (5 SIO campaigns to 15 non-SIO).}\label{tbl:F1}}
\end{table*}

\begin{figure}
\vspace{-0.1in}
\subfigure[Task 1, weekly, 1:3 SIO/non-SIO label ratio]{
	\includegraphics[width=0.9\columnwidth]{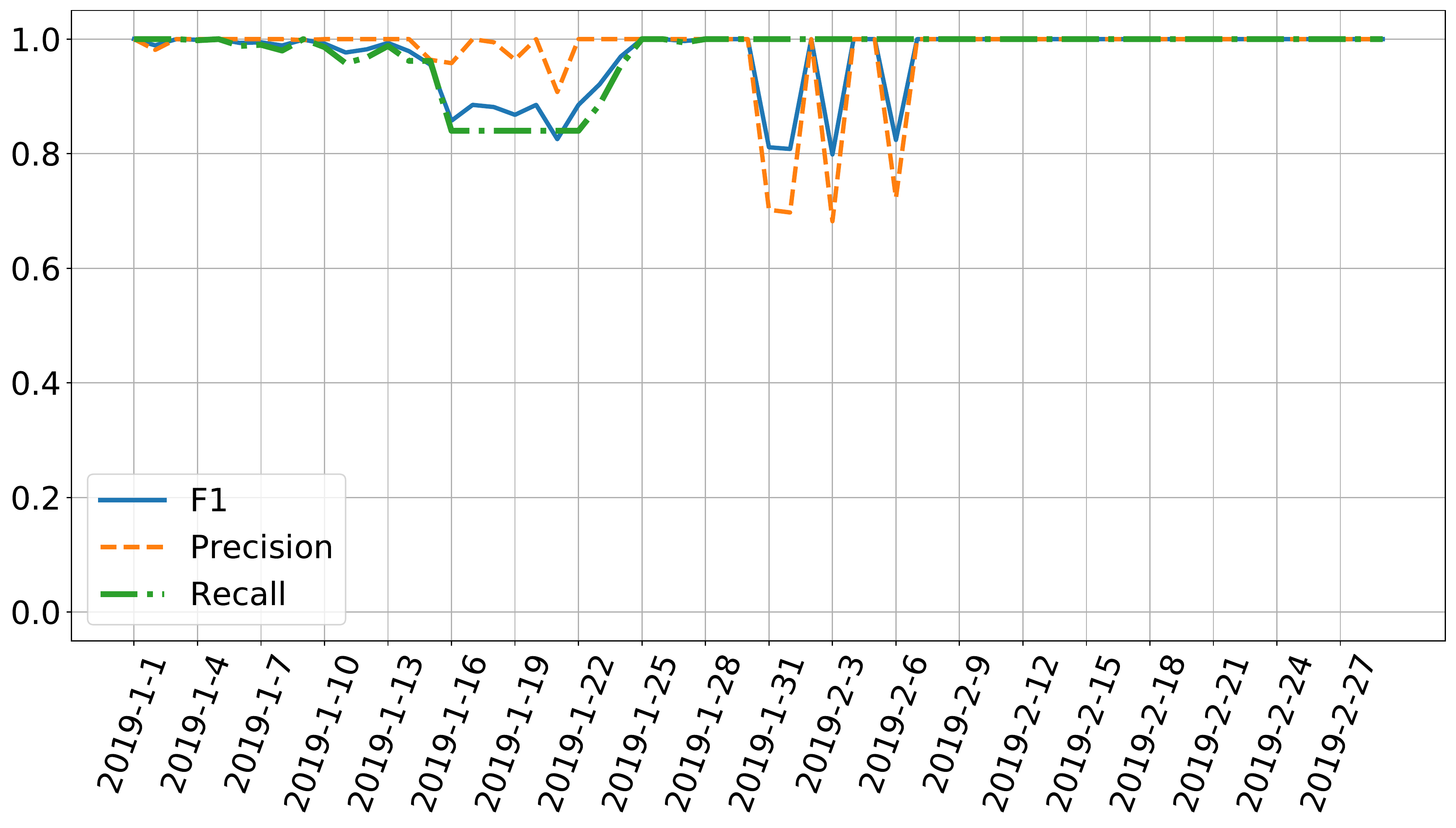}
	\label{fig:task1-weekly}
}
\subfigure[Task 2, weekly, 1:3 SIO/non-SIO label ratio]{
	\includegraphics[width=0.9\columnwidth]{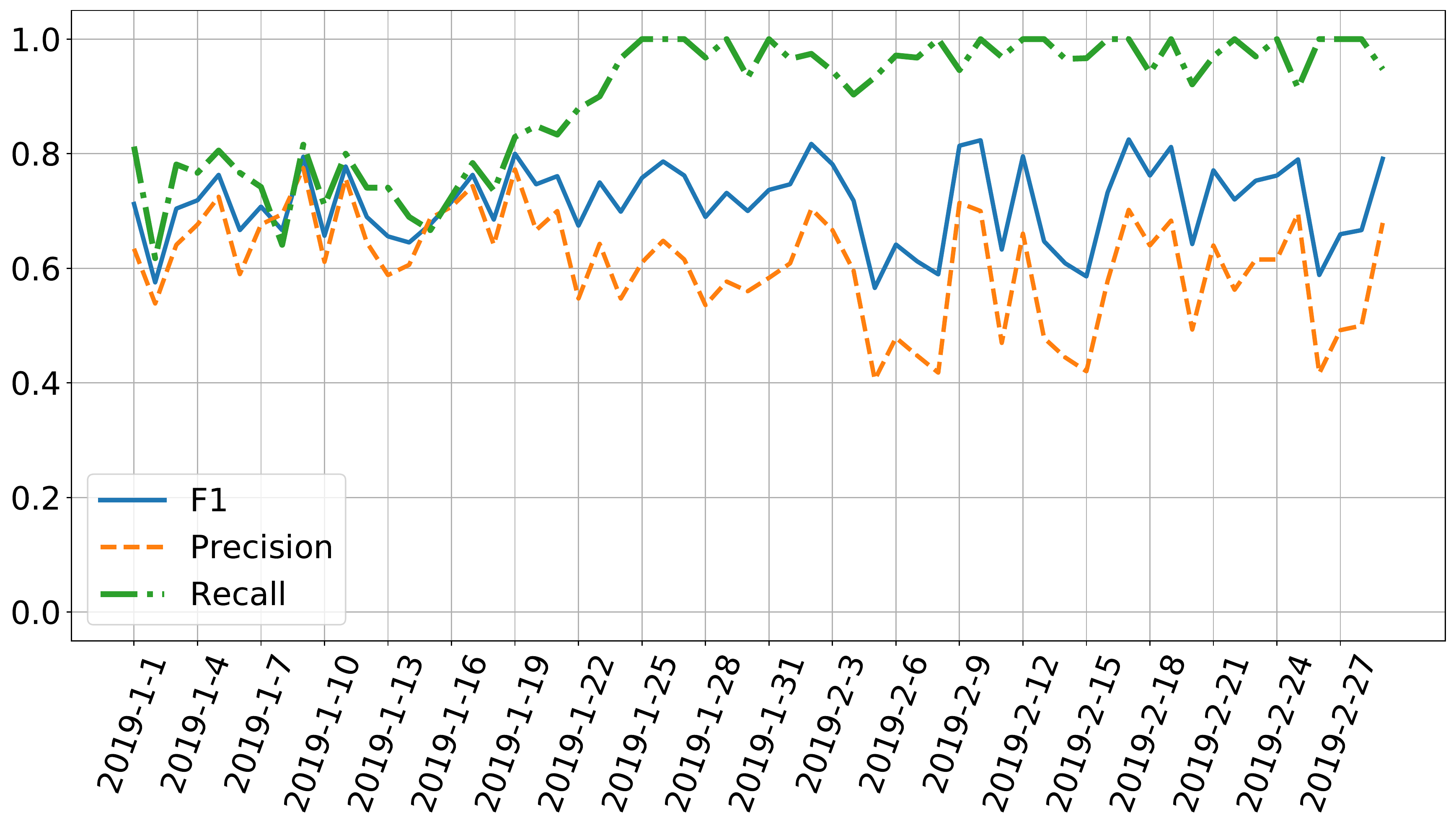}
	\label{fig:task2-weekly}
}
\vspace{-0.1in}
\caption{\newtext{Daily F1, precision, and recall scores from January 1st, 2019 to March 2nd, 2019. The 2019 US State of the Union address occurred on February 5th, 2019, which caused a surge in coordinated activity from US Congress Twitter accounts and confused our model. While an increase in the amount of coordinated activity from legitimate communities can lead to higher false positives (lower precision), it can also potentially help distinguish malicious from legitimate coordination (higher recall) depending on how different the legitimate coordination behavior is from those of SIO campaigns.}}
\end{figure}

\begin{figure}[t]
\includegraphics[width=0.9\columnwidth]{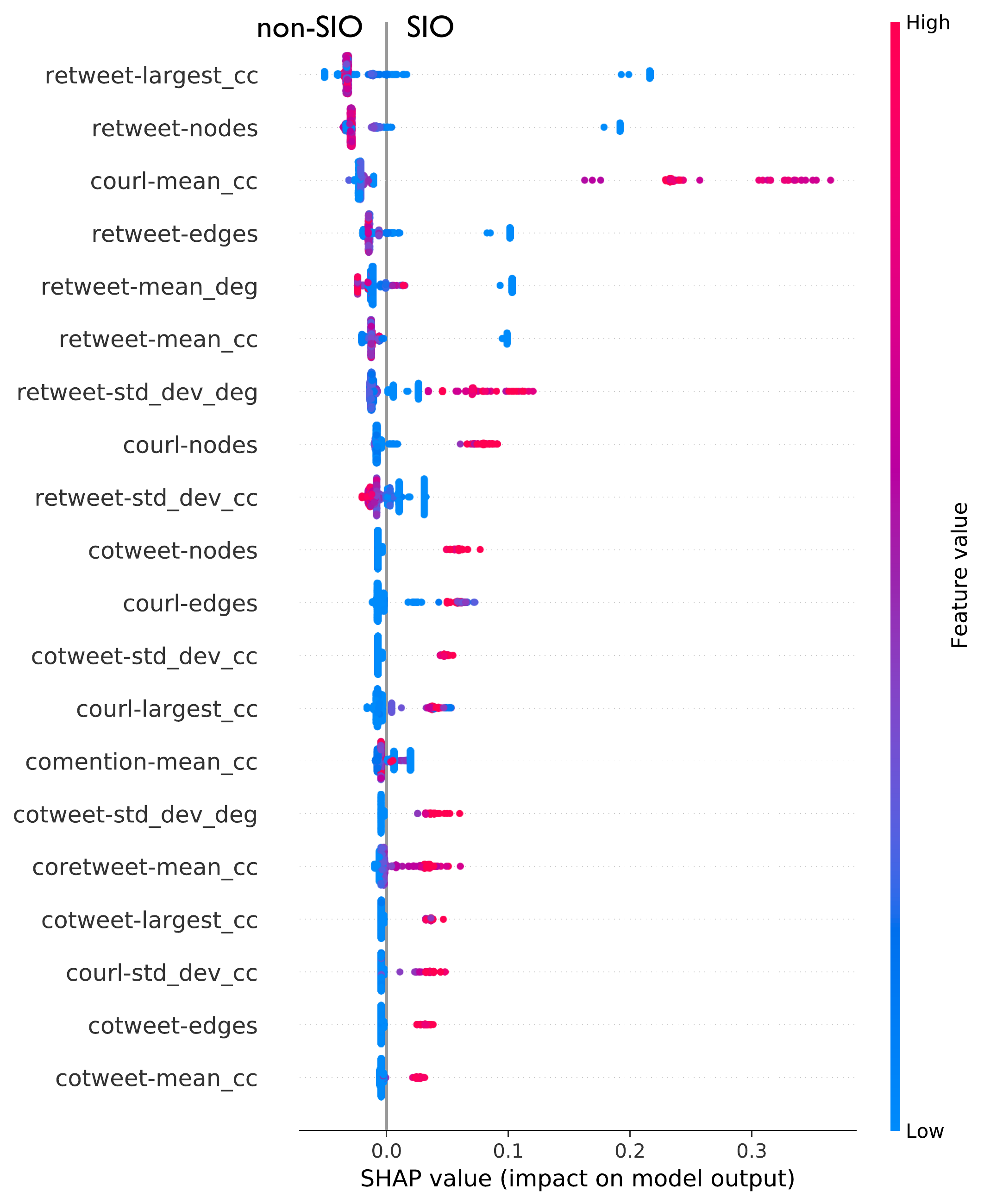}
\vspace{-0.2in}
\caption{SHAP feature importance (top to bottom). Retweet network activity metrics are most informative for the classifier to separate SIO campaigns and legitimate communities.}
	\label{fig:shapley}
\end{figure}

\newtext{
\subsubsection{Setup} For both \textit{Task 1} and \textit{Task 2}, we remove the first 60 days of the considered date range (April 1st, 2018---December 31st, 2019) when computing classification metrics to ensure there is no overlap with the tuning set. We average metrics over all subsequent days in the date range. For Task 1,  we compute 95\% confidence intervals for the 2:18 class ratio experiments over the 45 distinct ways to pair the SIO campaigns ${10 \choose 2}$, and over 100 randomly sampled sets of five campaigns for the 5:15 experiments. We compute the F1, precision, and recall scores for \textit{Task 2} over 120 prediction attempts ${10 \choose 3}$ for the 2:18 experiments and 100 prediction attempts for the 5:15 experiments. We compute 95\% confidence intervals for Task 2 by repeating each experiment with three distinct random seeds. Note that Task 2 has a different evaluation protocol than Task 1 because it requires having a held-out SIO campaign and Twitter community baseline, which in turn requires biased random sampling to achieve the proper class imbalance ratio. 


\subsubsection{Task 1 Results} 
The quantitative results for Task 1 are provided in Table~\ref{tbl:F1}. Overall, the RF achieves an F1 score of $0.980$ on this task. We observe a substantial improvement in recall when using weekly coordination instead of only daily (R = 0.853 $\rightarrow$ R = 0.971). The metrics for a 60-day range are visualized in Figure~\ref{fig:task1-weekly}. Although individual events, such as the US State of the Union address on February 5th, 2019, can lead to misclassifying legitimate behavior (low precision), we find that coordination network features are useful for successfully detecting near future activity exhibiting previously observed coordination patterns.

\subsubsection{Task 2 Results} 
The F1 scores for Task 2 (shown in Table~\ref{tbl:F1}) are notably lower than for Task 1. The best classification precision is P = 0.641 when using daily aggregation and a more balanced training setting. However, we see that the weekly aggregation improves recall from R = 0.732 to R = 0.877 with the advantage of using more SIO campaigns and a more balanced label ratio. This setup achieves the best F1 score of 0.709. The biggest challenge of classifying previously unseen SIO campaigns and legitimate activity at test time appears to be false positives. Indeed, we see a significant drop in precision from P = 0.985 in Task 1 to P = 0.641 in Task 2. A detailed look at the precision scores over a 60-day range is provided in Figure~\ref{fig:task2-weekly}.

\subsubsection{Model Interpretation and Further Analysis}
To further probe the classifier's decision-making process, we produced a SHAP summary plot of RF feature importance (Figure~\ref{fig:shapley}) by randomly selecting one day (March 2nd, 2019) from our training date range  and training a classifier for the  Task 2/weekly/5:15 label ratio setting. As our focus is primarily evaluating an out-of-the-box binary classifier using coordination network analysis features, we do not analyze the features in further detail by examining e.g., partial dependence plots or multicollinearity. We find that over the historical context for this target date, the classifier mostly relies on various statistics of the retweet network to separate SIO and non-SIO data. 

As shows in Figure~\ref{fig:shapley}, large values for the retweet network statistics appear to indicate \emph{non-SIO activity}, whereas large values for the co-url network's mean connected component is a primary indicator of SIO activity. While it may seem surprising that the largest connected component of the retweet network is a strong indicator for \emph{non-SIO} activity, recall that the retweet coordination pattern does not require a strict time threshold within which the retweet must occur. Over the span of a week, cohesive communities like members of the same political party or the same academic circles may (innocuously) retweet the same tweets, creating large and strongly connected retweet networks. Other important coordination features that strongly indicate similarity to SIO activity include the standard deviation of the retweet network node degree and the number of nodes in the co-url and co-tweet networks. Overall, stronger connectivity in the coordination networks---aside from the retweet network---increases the chances that the classifier labels the activity as SIO. 

}

\begin{figure}
	\includegraphics[width=0.49\textwidth]{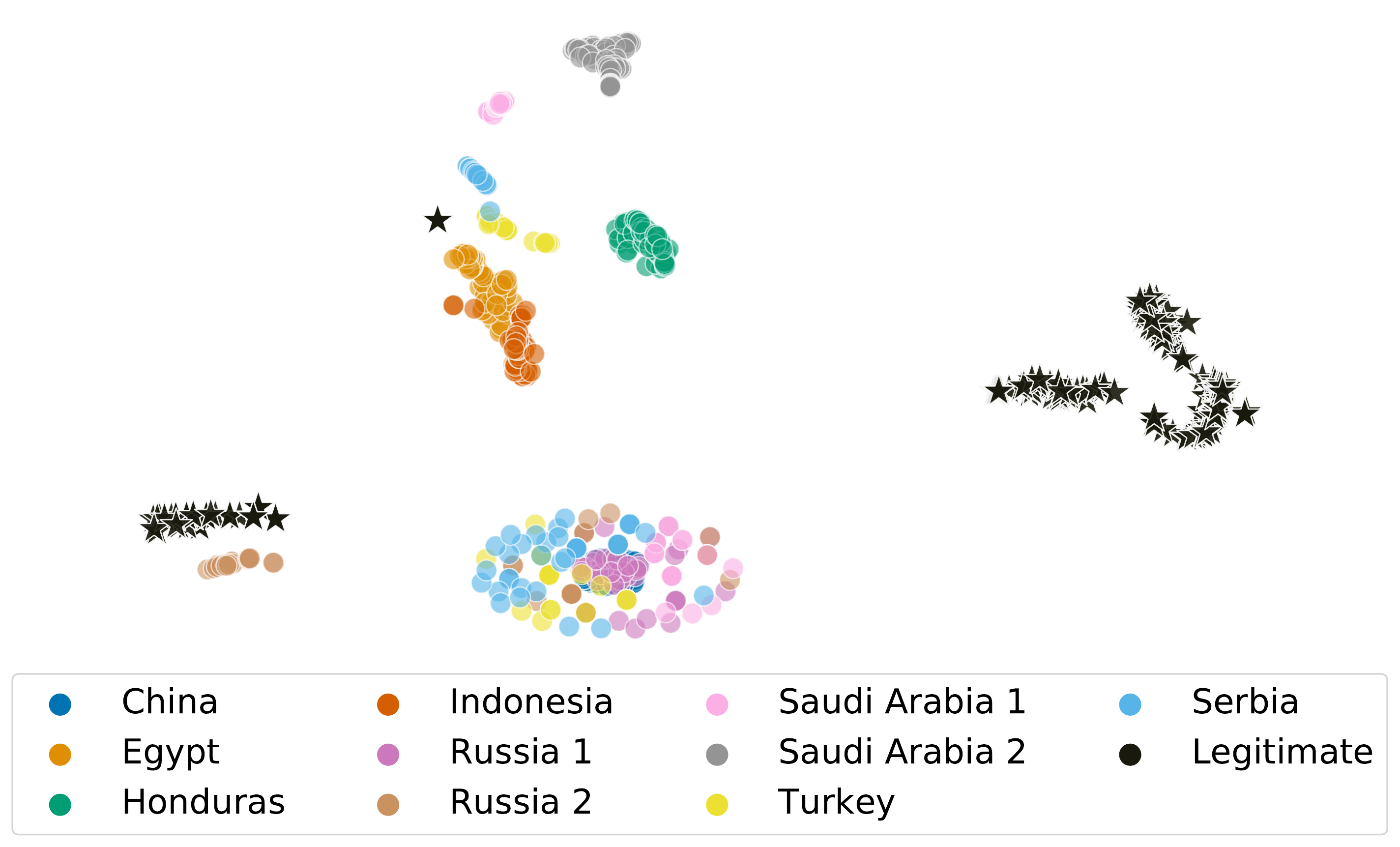}
	\caption{\newtext{T-SNE visualization of coordination activity across SIO campaigns in a 60-day time window (January 1st, 2019  to March 2nd, 2019). The tight cluster of overlapping campaigns at the bottom are days of little-to-no coordinated activity. Crucially, notice that the legitimate activity forms two distinct clusters with the smaller one appearing closer to other SIO activity, thereby contributing to a higher false-positive rate. In general, SIO coordinated activity appears to cluster by campaign identity with minimal overlapping in between. The minimal overlap highlights the difficulty of Task 2, indicating that coordination patterns exhibited by one group of campaigns are not usually informative of others co-occurring in the same time (e.g., Honduras vs. Russia 2). }
	}
	\label{fig:tsne}
\end{figure}

\newtext{
We also visualize a t-SNE~\cite{maaten2008visualizing} plot of the weekly-aggregated features over 60 days (Figure~\ref{fig:tsne}). We found that PCA~\cite{tipping1999probabilistic} was unable to produce an interpretable representation of coordination activity in our data; t-SNE is a more expressive dimensionality reduction tool for visualizing high dimensional data in a 2-D plane. We color the features by campaign source to help reveal similarities across distinct campaigns. First, days of low coordination activity manifest in the plot as the tight, multi-colored cluster at the bottom. Note that the classifier can use the number of days of low activity within the historical window as an additional signal. Second, the Twitter community baselines are split across two tight clusters. The smaller of the two clusters is situated close to SIO activity and represents days of increased coordination (e.g., because of a political event that the US Congress and UK Parliament respond to). This can confuse the classifier into believing the legitimate coordination is malicious. Third, and most importantly, the SIO campaigns appear to be tightly clustered but have minimal overlap. This helps explain the difficulty of Task 2; the coordination tactics of one set of SIO campaigns are not guaranteed to be highly informative for detecting other campaigns. Moreover, few patterns may exist that are shared across many campaigns, which can lead the model to underfit the data and extract very simplistic rules for identifying out-of-distribution SIO activity.
}

\subsubsection{Takeaways}

\newtext{
We conclude that the RF classifier is a promising step towards SIO activity detection but needs further improvement before deployment. These results show that methods based on coordination network analysis must balance effectiveness for usability. With a loose restriction on the number of allowable false positives, a classifier can identify out-of-distribution SIO activity (R = 0.877), but such a system would be impractical as it would flag a high amount of benign activity.  Therefore, while the RF classifier cannot \emph{currently} distinguish daily/weekly activity as SIO or non-SIO at desirable levels for \textit{all} Twitter communities, we believe these initial results can guide more sophisticated techniques.

}

\subsection{Examples of Misclassifications}
\label{sec:analysis:misclass}

\subsubsection{Setup} 
\newtext{Since there is a stark drop in classifier precision for Task 2 (meaning an increase in false positives), we now turn our attention to understanding why legitimate communities' coordinated activity gets misclassified as SIO activity. Using the weekly/5:15 experiment, we aggregate model predictions over all days across the 100 runs and examine the histogram of daily errors. We identified the ten days where the held-out non-SIO activity was most often misclassified across the 100 runs and } extracted all information from the tweets used to generate the coordination networks from each baseline community. Finally, we manually looked at both the tweets and the news for that specific date that may have affected the community's coordination patterns.

\subsubsection{Key Results} 
\newtext{
First, we observe that the histogram of daily errors is skewed right; essentially, across the 100 runs, there were a few days that were consistently misclassified, while a large fraction of days were not consistently misclassified across each of the runs. However, the majority of misclassifications in any given run comes from the latter set. Out of the ten most frequently misclassified days that we examined, we attributed three of them to a community reacting to a specific event that caused accounts to exhibit SIO-like coordinated activity. Those three events were the US Congress reacting to the 2019 State of the Union Address, US Congress reacting to the president's national emergency declaration in February 2019, and the UK Parliament reacting to the 2019 Remembrance Sunday holiday. All three of these days had significant activity in the retweet and hashtag coordinated networks. For the other seven days, we could not attribute a specific event that caused the misclassification. These days were hallmarked by relatively low amounts of coordinated activity consisting of a mixture of various coordination types (beyond retweeting). 
}

\subsubsection{Takeaways}
\newtext{
The findings above show us that further improvements in coordinated activity detection requires a strong foundational understanding of what legitimate coordinated activity looks like on Twitter. The prescence of a mixture of coordination types on the misclassified days suggests that the classifier struggles to learn sophisticated patterns for legitimate coordination that generalize to unseen activity. This makes the classifier sensitive to the training data it sees, and it often seems to rely on simplistic rules based on the most commonly seen legitimate coordination pattern: retweeting (recall the discussion about the SHAP plot results in Section~\ref{sec:analysis:classification}).  Improving our knowledge of how legitimate communities coordinate will help these models extract better patterns that discriminate SIO coordinated activity from benign behavior. Additionally, we found that false positives in coordinated activity detection also happen when the online world reacts to real-world events. Even though each event may only affect one community, this phenomenon is expected to be a common occurrence across Twitter.
}

\section{Discussion}
\label{sec:disc}

\subsection{Limitations}
\label{sec:disc:limitations}

While the SIO coordination patterns examined have been used for astroturfing detection and forensic analysis~\cite{ratkiewicz2011detecting,keller2020political}, and to some extent represent normal behavior of Twitter users, other coordination patterns likely exist. For example, a common behavior that may be heavily used by Twitter communities---but is not included in our analysis---is sharing the same image or video. We also note that our ground truth data for SIO activity is based on {\it discovered} SIO campaigns Twitter has published. Likely, many more SIO campaigns are still operating on the Twitter platform. Some of the coordination patterns present in those communities are likely unaccounted for by our analysis.

\subsection{Coordination as a Spectrum}
\label{sec:disc:spectrum}

As we showed in our experiments, coordination is not a unique phenomenon that only occurs in SIO campaigns. Each of the four collected baselines shows varying levels of coordinated activity, with political baselines exhibiting both SIO-like and non-SIO coordination patterns. Since other benign communities (e.g., governments, political activists) likely share similar behaviors as our political baselines, it is essential to note that their {\it SIO coordination activity should be seen as a spectrum and not a binary state.} Forcing a classifier to make a binary decision between SIO and non-SIO based on coordination can lead to an overestimation of accounts that are part of disinformation campaigns. These overestimations would flag the activity of legitimate accounts as suspicious and possibly lock or suspend them, thereby degrading the usability of Twitter. 

\subsection{Detection Outside the Closed World}

Current machine learning techniques make a strong assumption that the future will resemble the past. In adversarial environments that are constantly changing, this assumption rarely ever holds. Events that are inherently unpredictable pose significant challenges for even anomaly-detection systems, which are trained to identify abnormal data points, as examples of the ``novel'' data points do not yet exist at training time~\cite{sommer2010outside}. \newtext{The coordination tactics employed by one SIO campaign may look totally different from the tactics used by the next uncovered campaign.} This problem is akin to the ``hindsight is 20/20'' expression which tells us that decisions made in the past are easy to understand once we look back at them but hard to justify as they are happening. For these reasons, we emphasize that research into SIO detection must prioritize gaining \textit{insight} over improving the numerical results.

\subsection{Future Work}
\label{sec:disc:future}

In this work, we focus on separating the activity of Twitter communities, whether SIO campaigns or legitimate users, based on their daily and weekly coordination patterns. While these time periods give us insight into how coordinated a community is for a specific day and week, they provide little information as to how activity changes over time. Our current method relies on sliding window time series prediction which alone provides only a weak temporal signal. Future research in this field could expand into finding ways to incorporate temporal dependencies into SIO coordination classifiers\newtext{, e.g. with recurrent neural networks~\cite{mazza2019rtbust}}. By looking at the evolution of coordination, more behavior patterns of SIO campaigns and Twitter communities may emerge, such as cycling of accounts or measuring the burstiness or continuous coordination fingerprint of a community. \newtext{Another important future direction is to study how human operators can best leverage automated tools for inauthentic coordinated activity detection to rapidly identify SIO campaigns.} Finally, as evidenced by the analysis in Section~\ref{sec:analysis:misclass}, collecting more organic and legitimate Twitter community activity would be beneficial for fitting more powerful classifiers.

\section{Conclusion}
\label{sec:conc}

Strategic Information Operations have exploited online social media sources to deceive and manipulate online rhetoric. Timely detection of these SIO campaigns, however, is still an open problem. \newtext{While previous works concentrate on detection at an account level, we focus on distinguishing an information operation's group coordinated activity from other normal behavior in Twitter to provide evidence for human analysts to do further investigations. To do this, we explored coordination networks used previously for forensic analysis to represent a Twitter community's activity. As our results show, known patterns in coordinated activity can be used to reliably distinguish SIO-like activity from the regular coordinated activity found on Twitter. While this is a step in the right direction, more work is needed in coordinated activity classification as previously unseen SIO campaigns and legitimate community activity decreases classification performance at test time.
} 


\newtext{
\section*{Acknowledgements}
\label{sec:ack}

The authors would like to thank our anonymous reviewers for their helpful comments.
We would also like to thank Ali Sadeghian for our numerous discussion. 
This work was supported by the National Science Foundation grant numbers CNS-1562485. Any findings, comments, conclusion found in this work are from the authors and may not necessarily reflect the views of the funding agency.

}

\bibliographystyle{ACM-Reference-Format}
\bibliography{disinfo}

\clearpage

\appendix
\section{Appendix}
\label{appendix:mapping}

\begin{figure*} 

	\subfigure[Unknown $\rightarrow$ Serbia]{
		\includegraphics[width=0.235\textwidth]{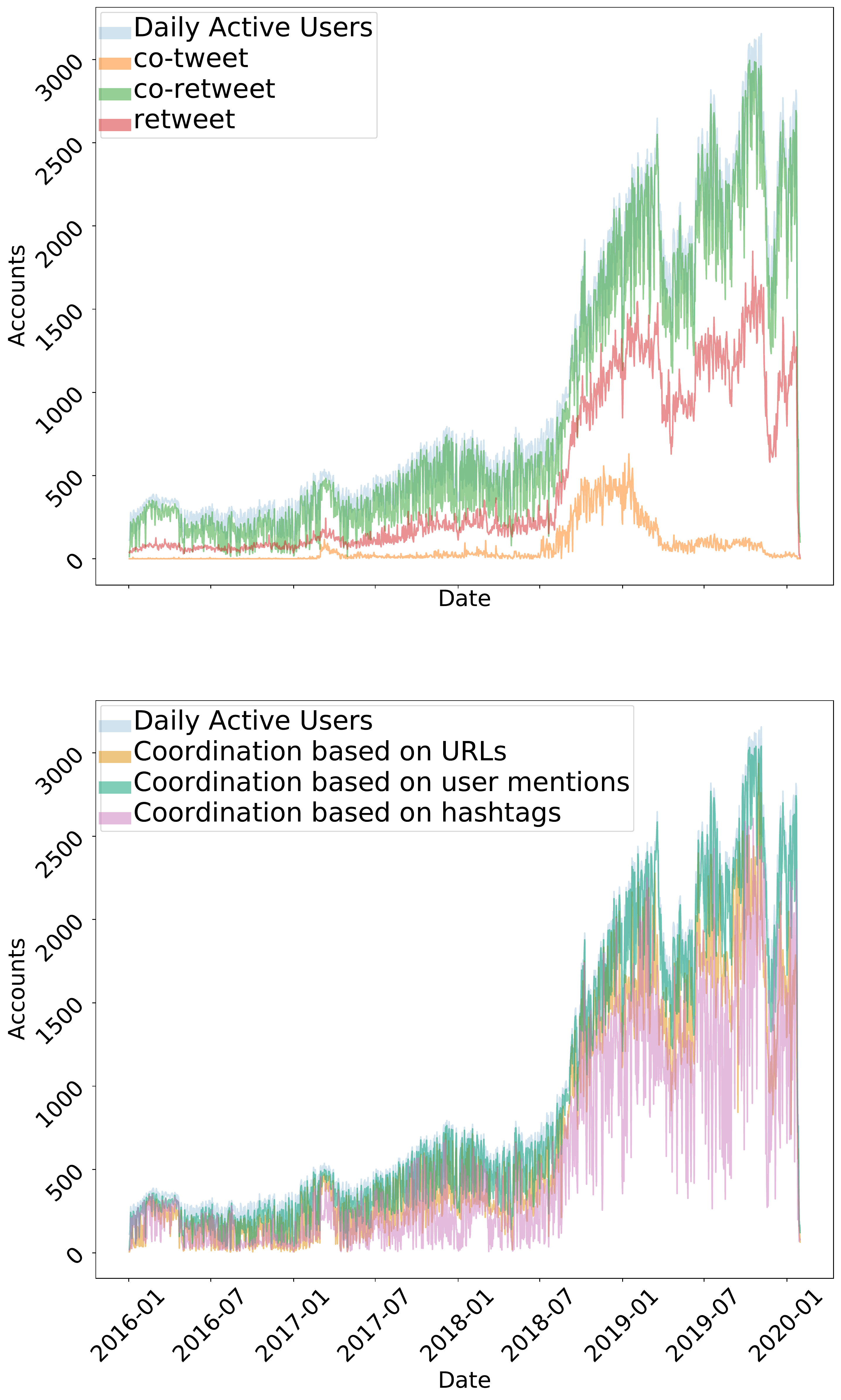}
		\label{fig:catalonia}}
	\subfigure[Saudi Arabia (external) $\rightarrow$ Saudi Arabia]{
		\includegraphics[width=0.235\textwidth]{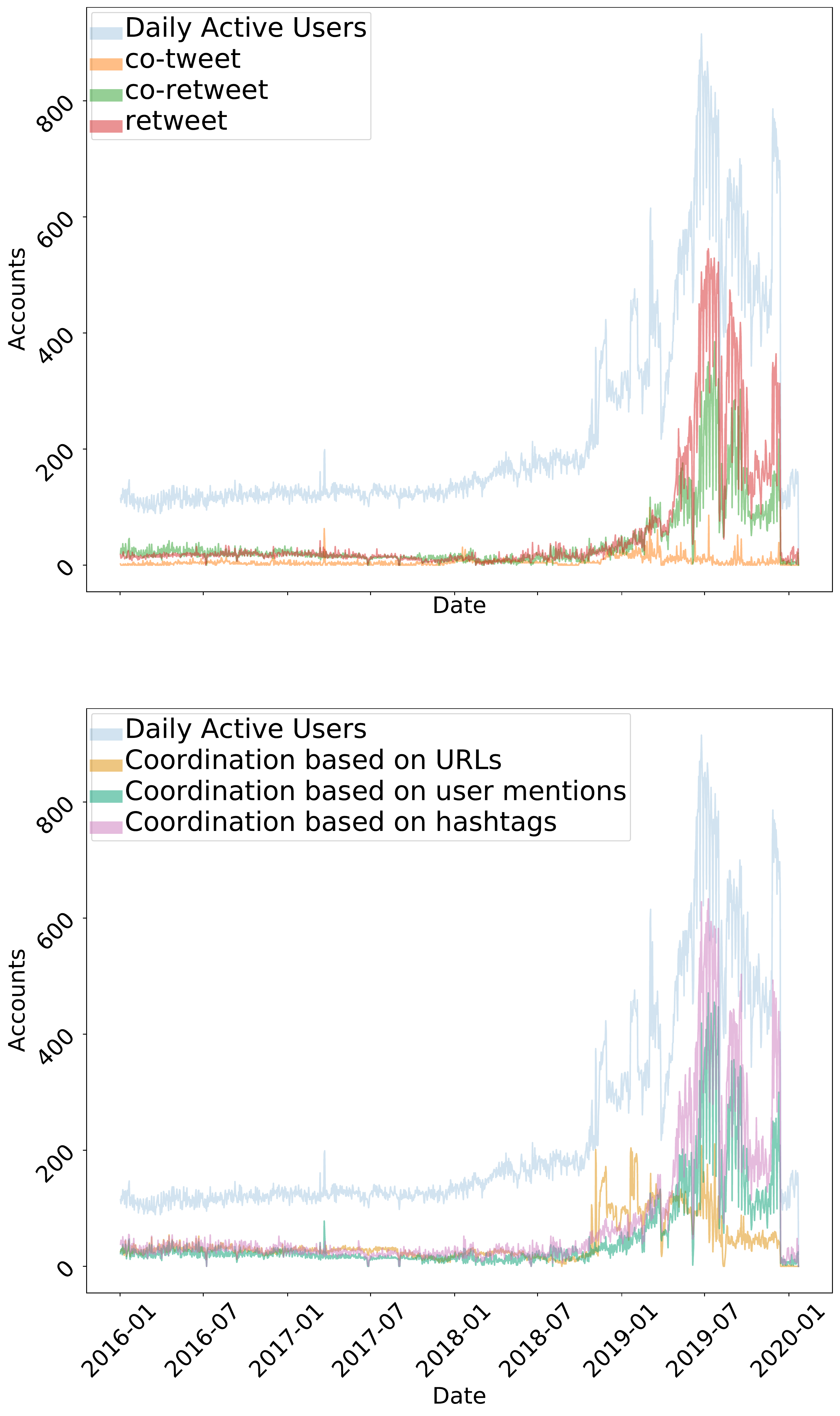}
		\label{fig:china}}
	\subfigure[Egypt $\rightarrow$ Iran/Quatar/Turkey]{
		\includegraphics[width=0.235\textwidth]{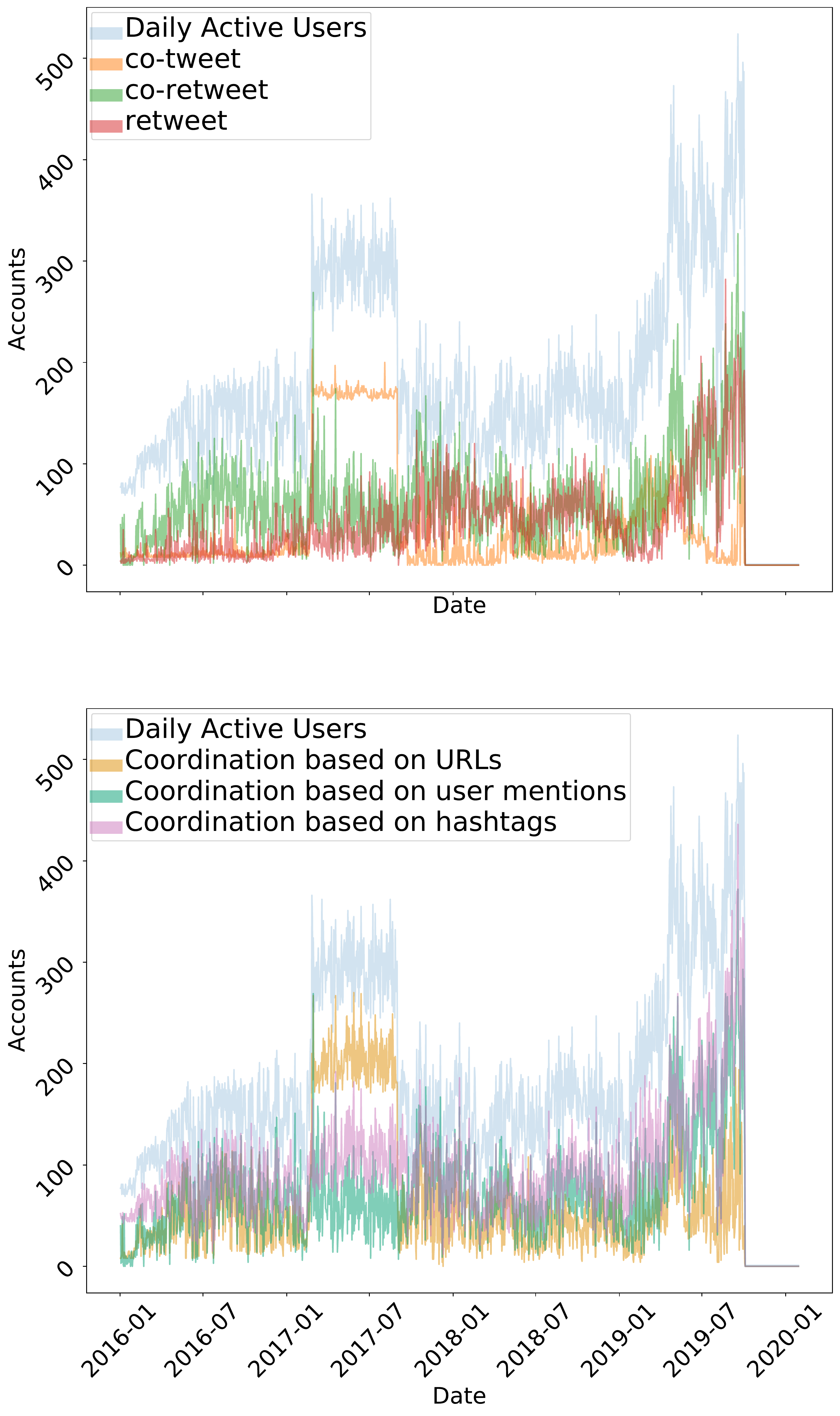}
		\label{fig:ecuador}}
	\subfigure[Russia (internal) $\rightarrow$ Russia]{
		\includegraphics[width=0.235\textwidth]{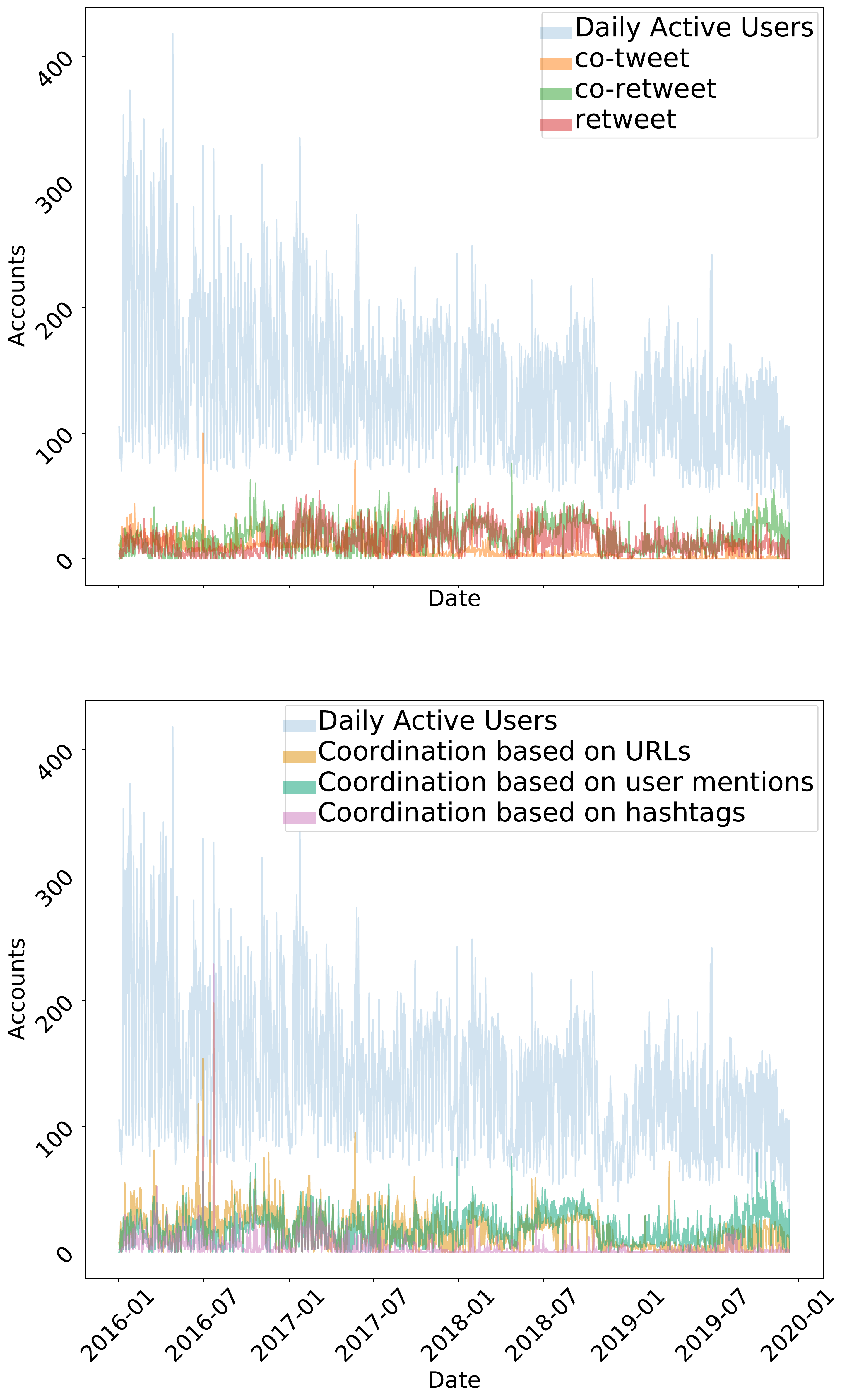}
		\label{fig:ira}}

	\subfigure[Indonesia $\rightarrow$ Indonesia]{
		\includegraphics[width=0.235\textwidth]{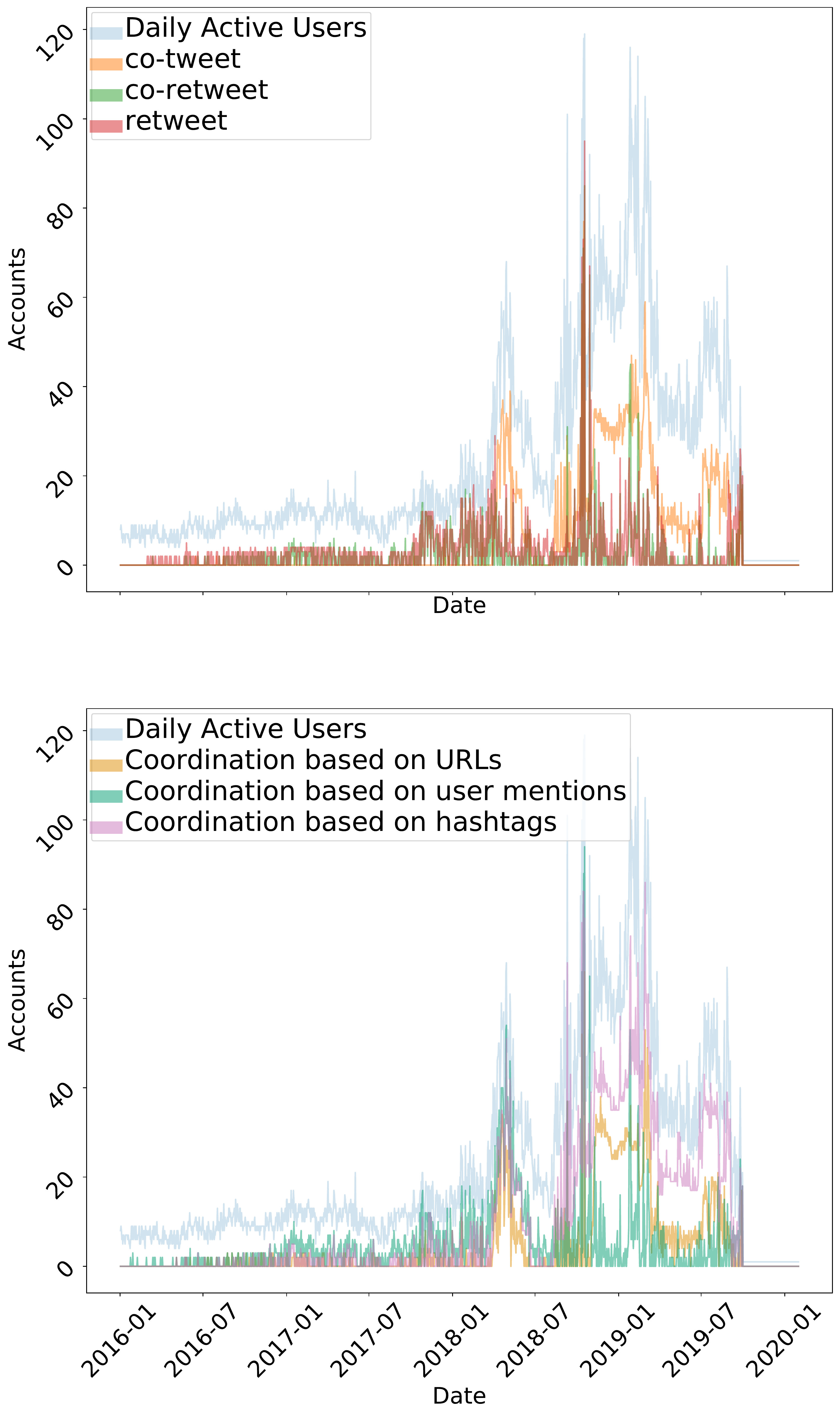}
		\label{fig:uae_iran}}
	\subfigure[Honduras $\rightarrow$ Honduras]{
		\includegraphics[width=0.235\textwidth]{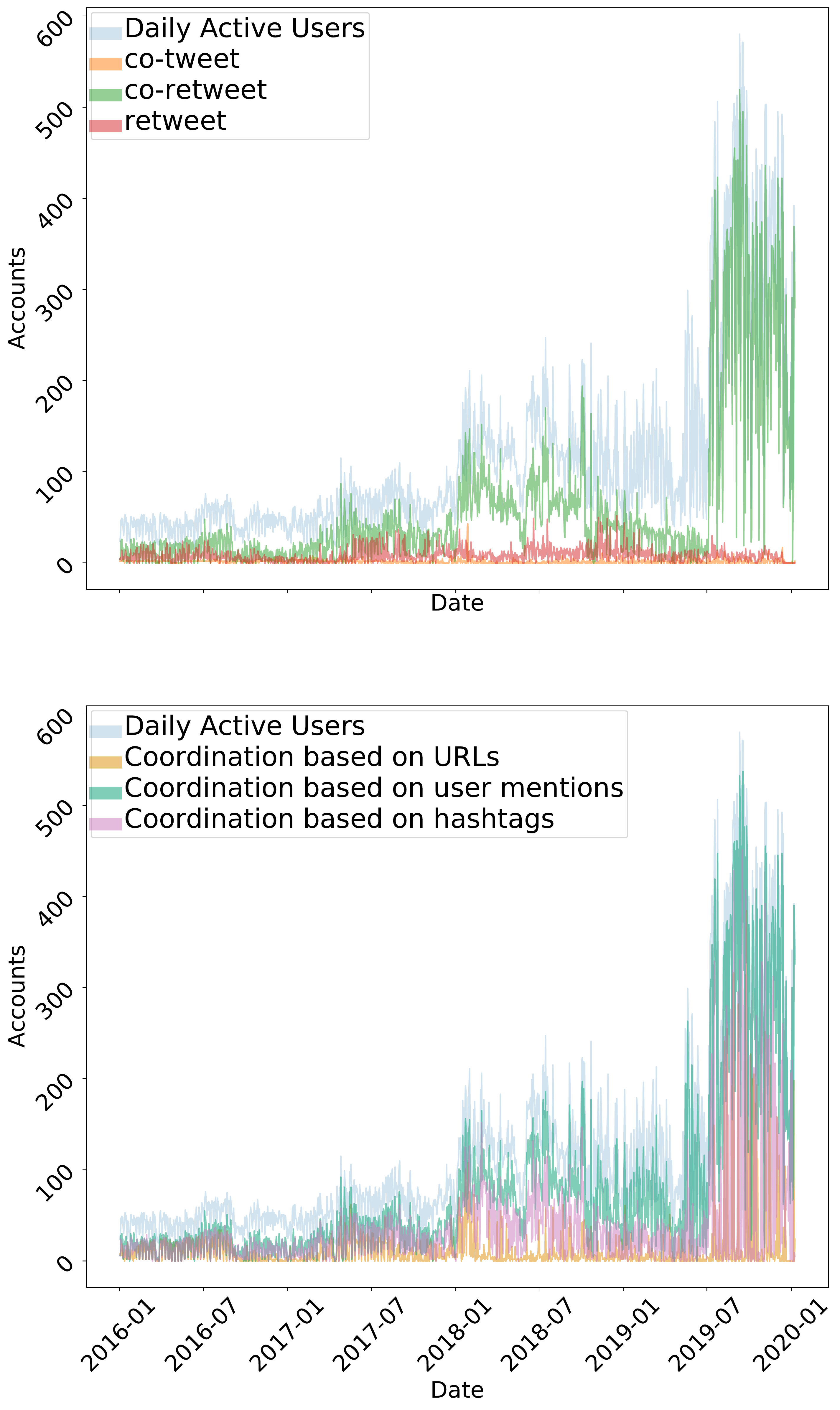}
		\label{fig:iran_usa}}
	\subfigure[China (2020) $\rightarrow$ Hong Kong]{
		\includegraphics[width=0.235\textwidth]{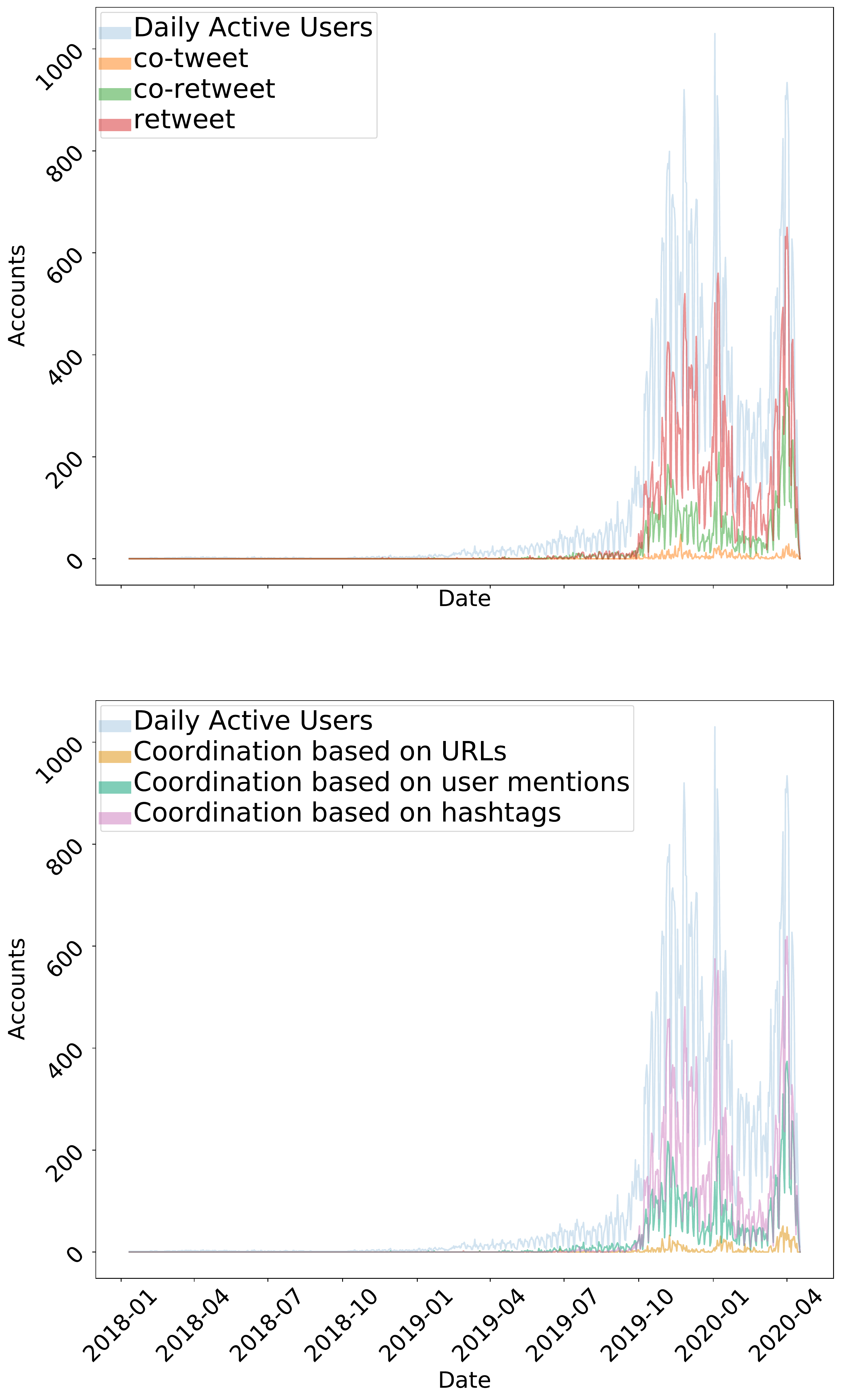}
		\label{fig:venezuela_venezuela}}
	\subfigure[Russia (external) $\rightarrow$ Russia]{
		\includegraphics[width=0.235\textwidth]{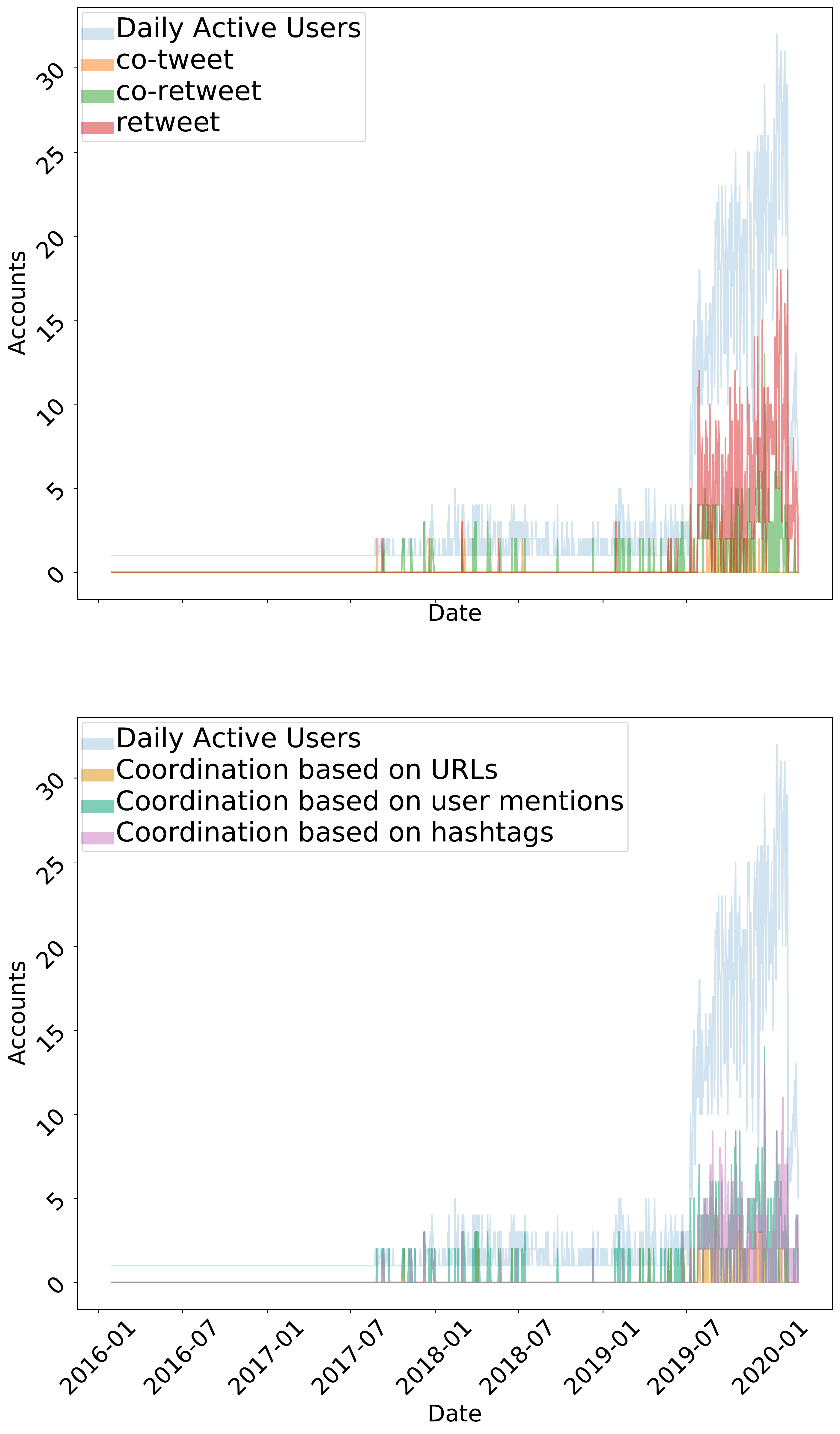}
		\label{fig:uae_quatar_yemen}}

	\subfigure[Parliament (Baseline)]{
		\includegraphics[width=0.235\textwidth]{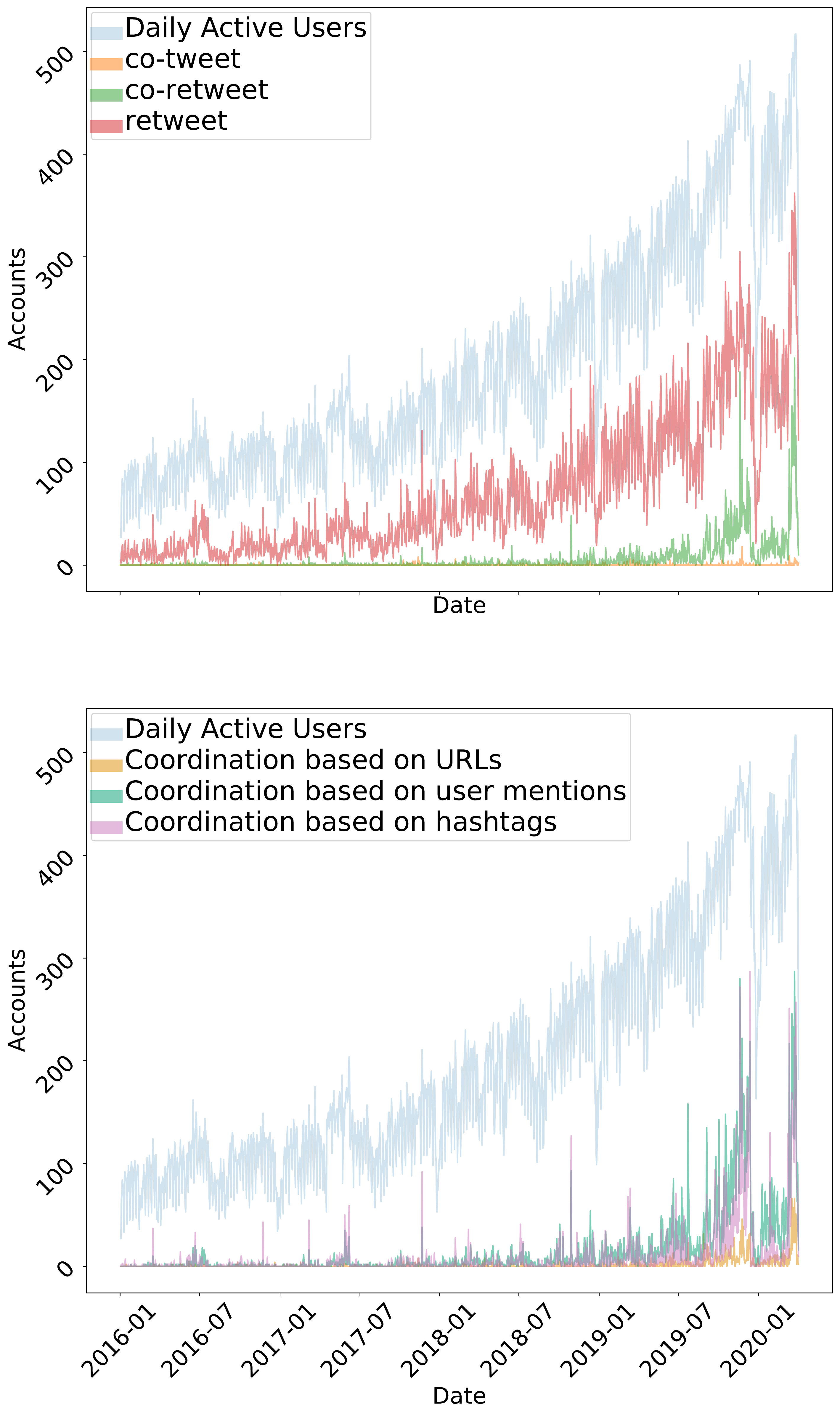}
		\label{fig:parliment}}
	\subfigure[Random (Baseline)]{
		\includegraphics[width=0.235\textwidth]{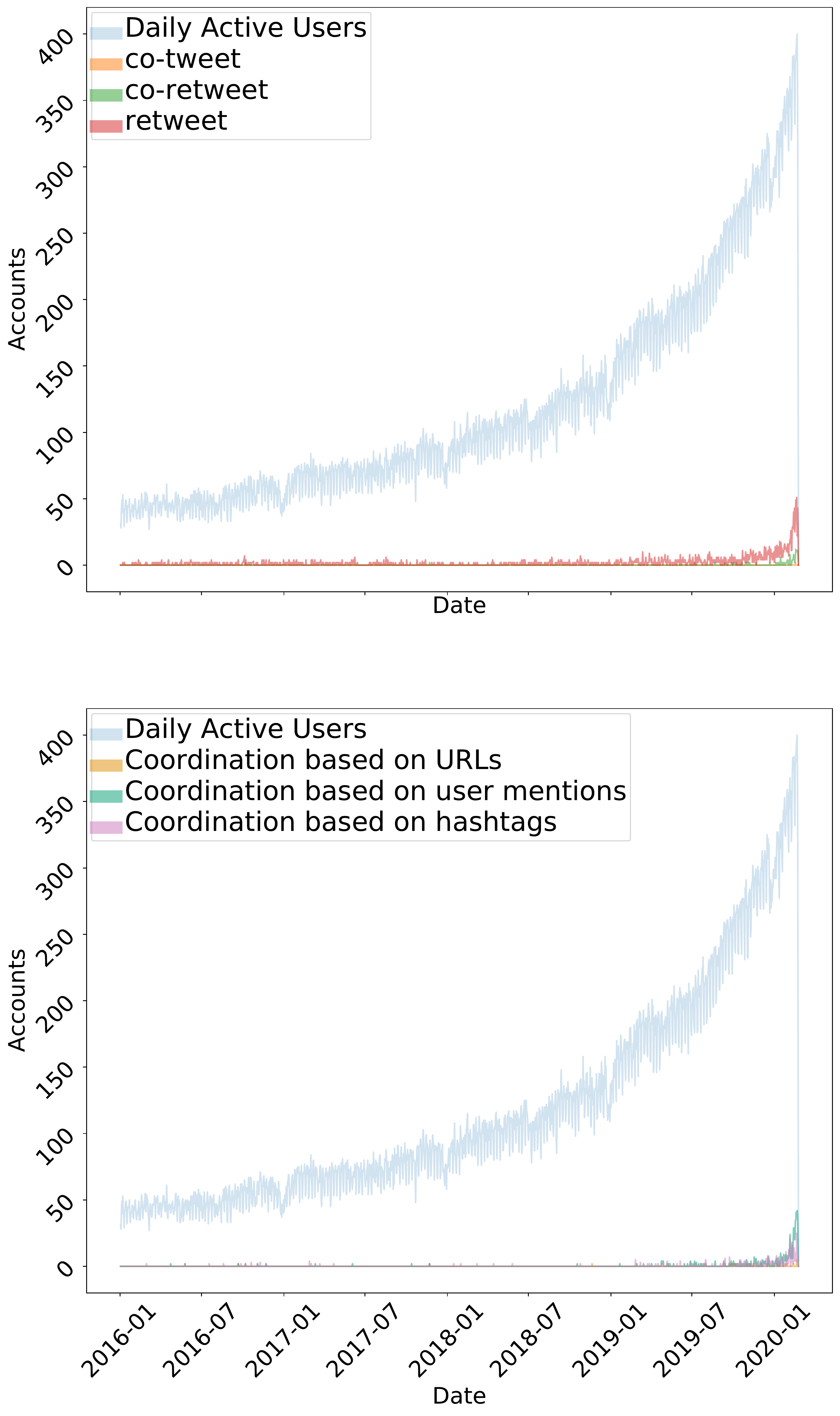}
		\label{fig:random}}
	\subfigure[Academics (Baseline)]{
		\includegraphics[width=0.235\textwidth]{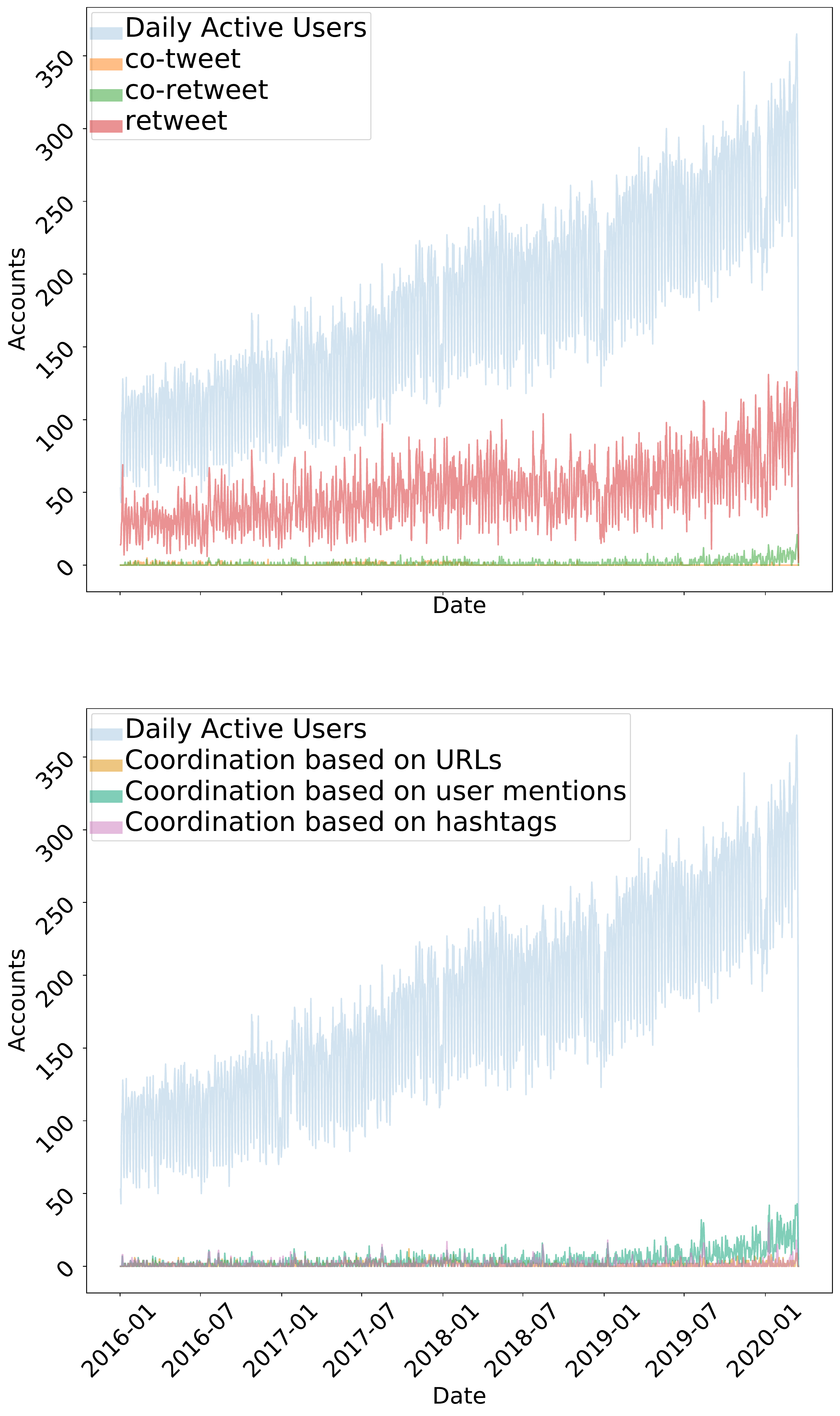}
		\label{fig:academics}}

	\caption{This Figure shows the coordination patterns of the remaining eight disinformation campaigns and the three community baselines not shown in Section~\ref{sec:analysis:results}. Note the y-axis are not the same values for different campaigns.}
	\label{fig:appendix_activity}
\end{figure*}

\begin{figure*}
	\subfigure[P-R curve]{
		\includegraphics[width=0.45\textwidth]{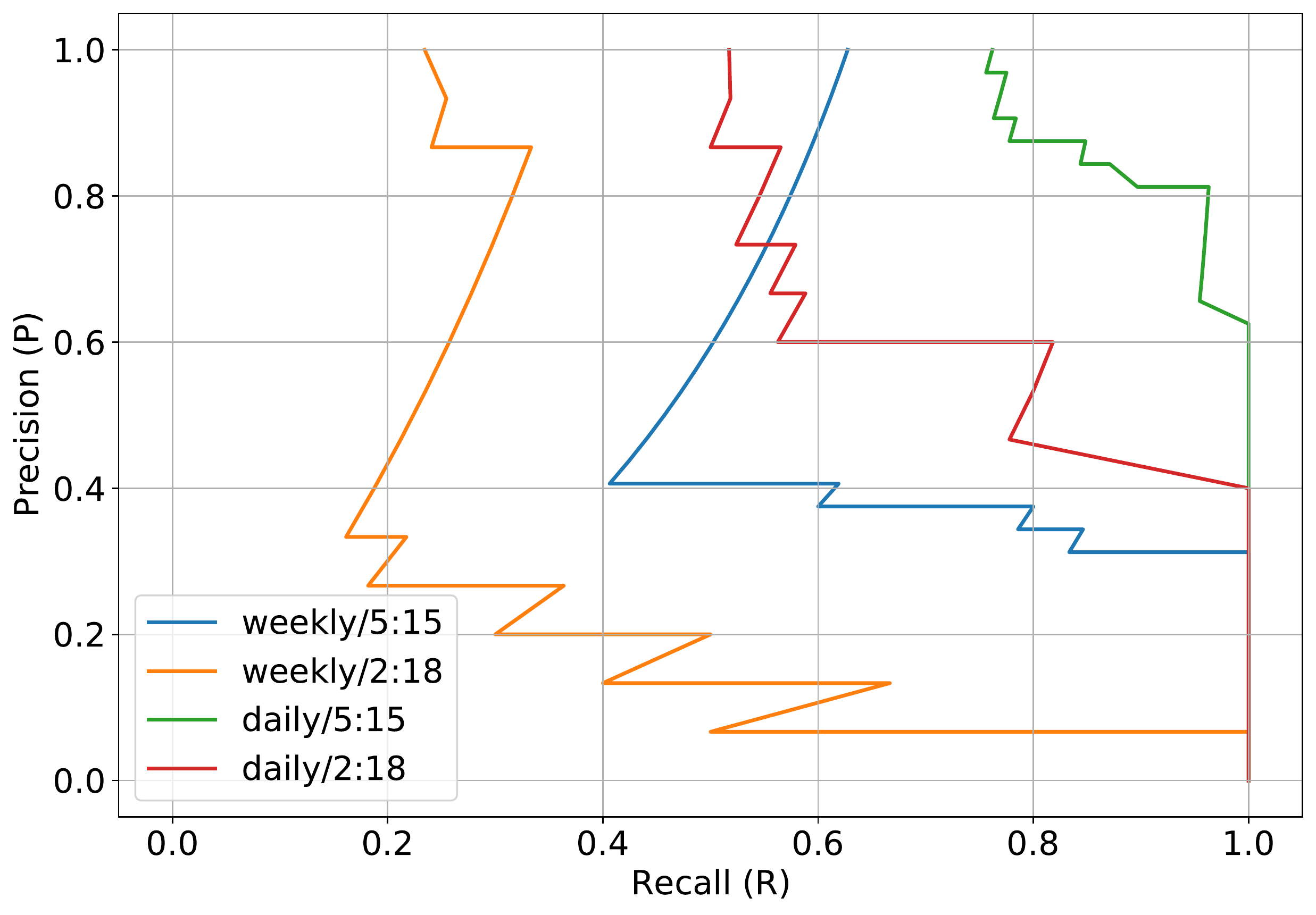}}
	\subfigure[P-R curve]{
		\includegraphics[width=0.45\textwidth]{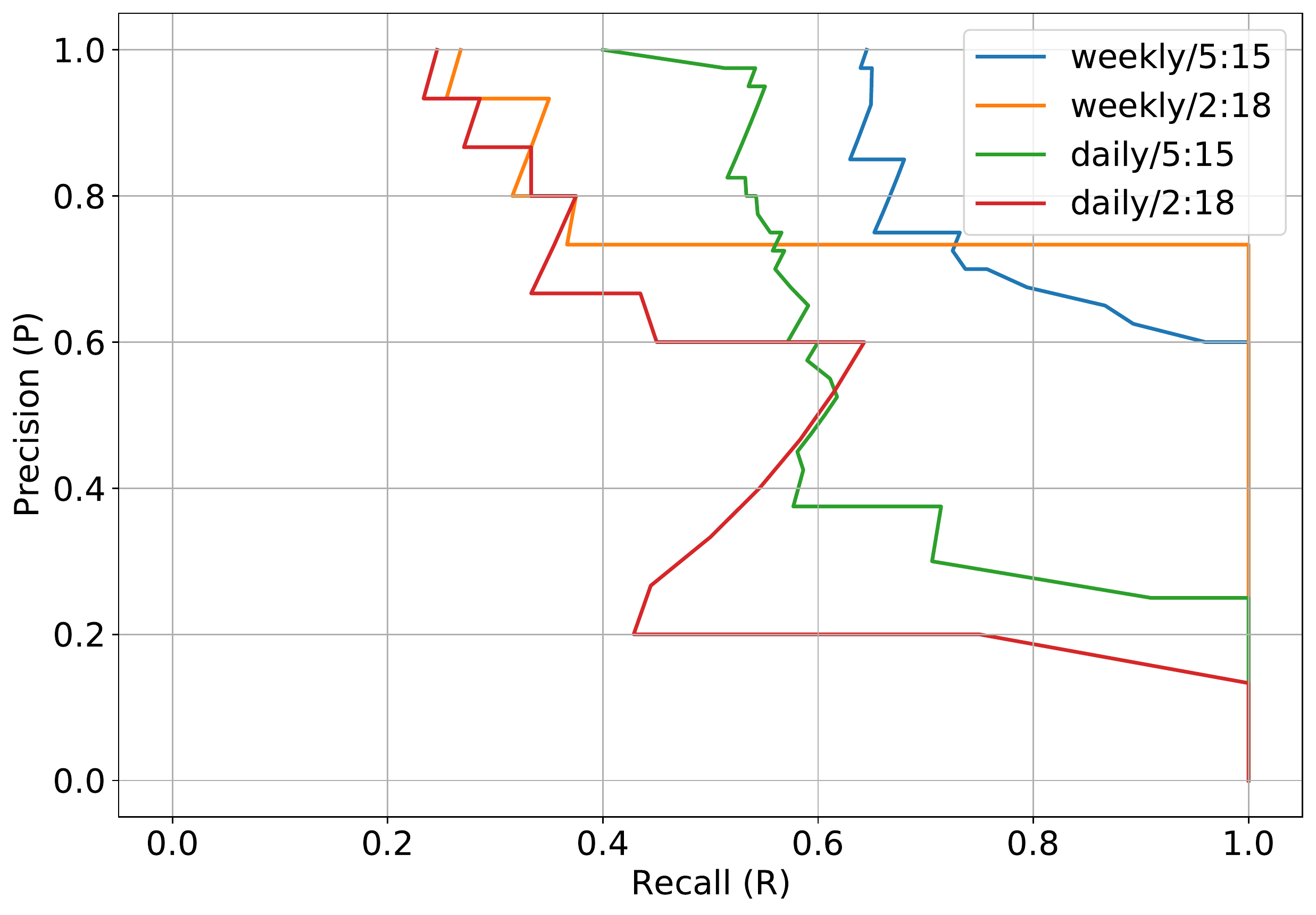}}
	\subfigure[P-R curve]{
		\includegraphics[width=0.45\textwidth]{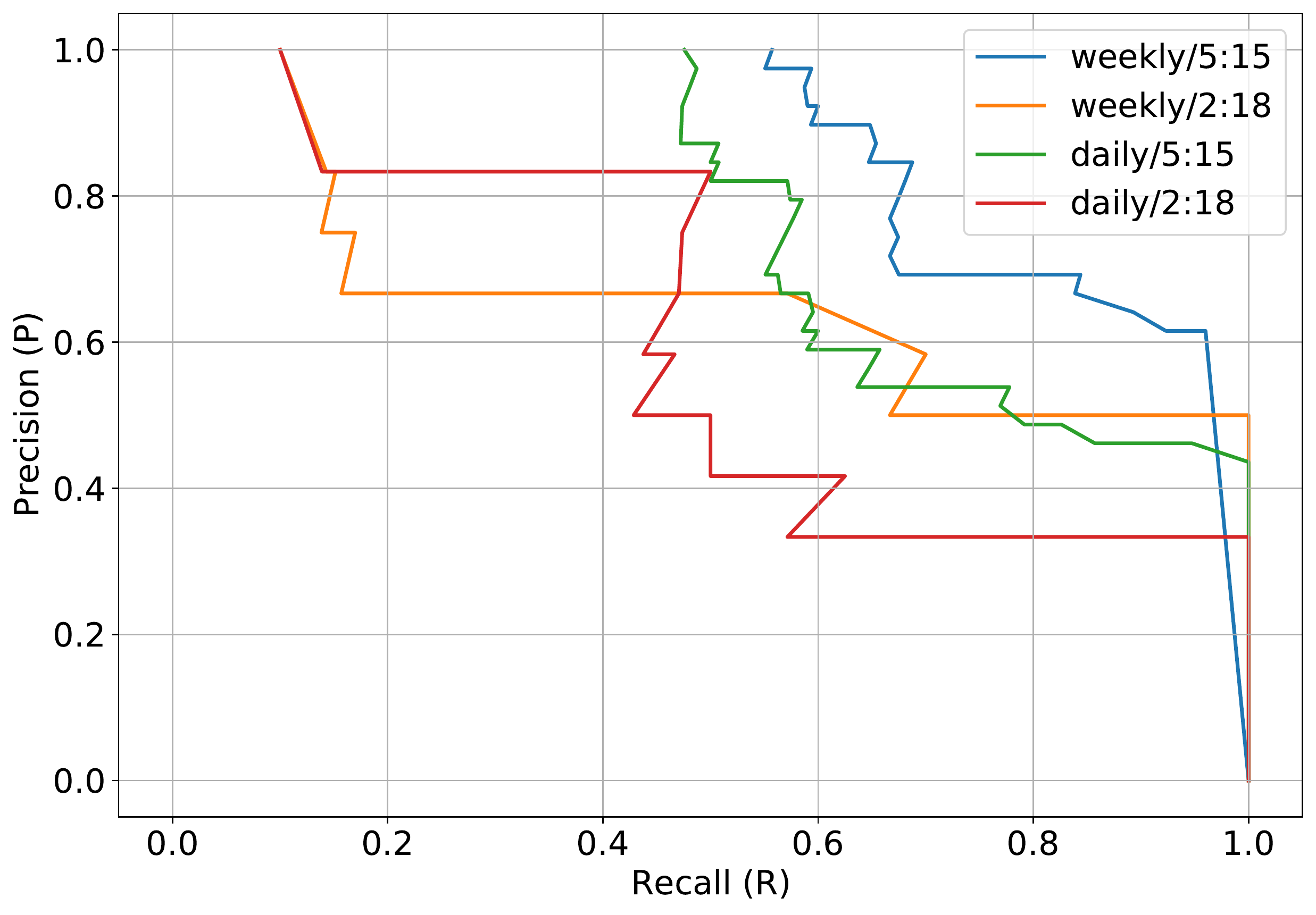}}
	\subfigure[ROC curve]{
		\includegraphics[width=0.45\textwidth]{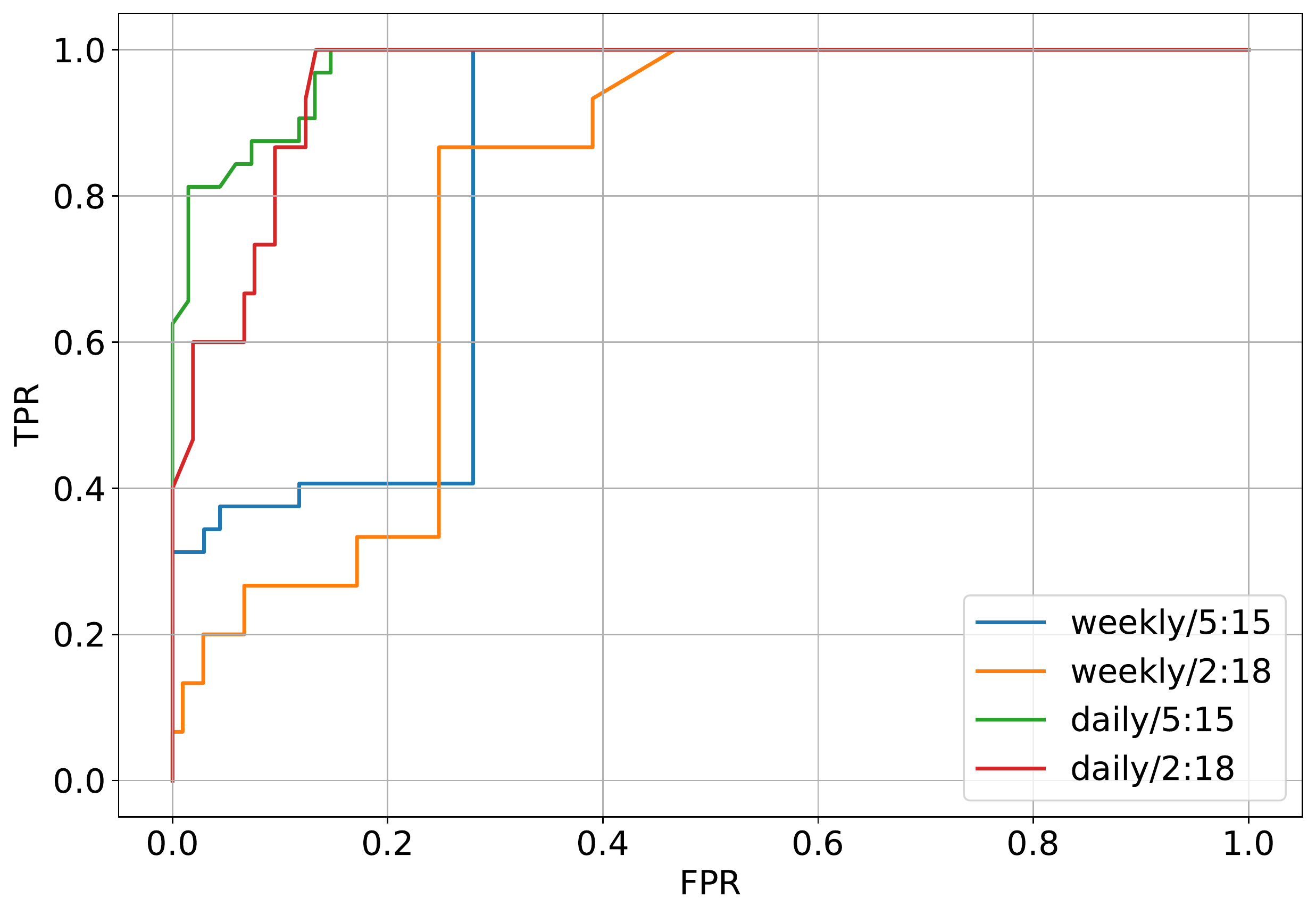}}		
	\subfigure[ROC curve]{
		\includegraphics[width=0.45\textwidth]{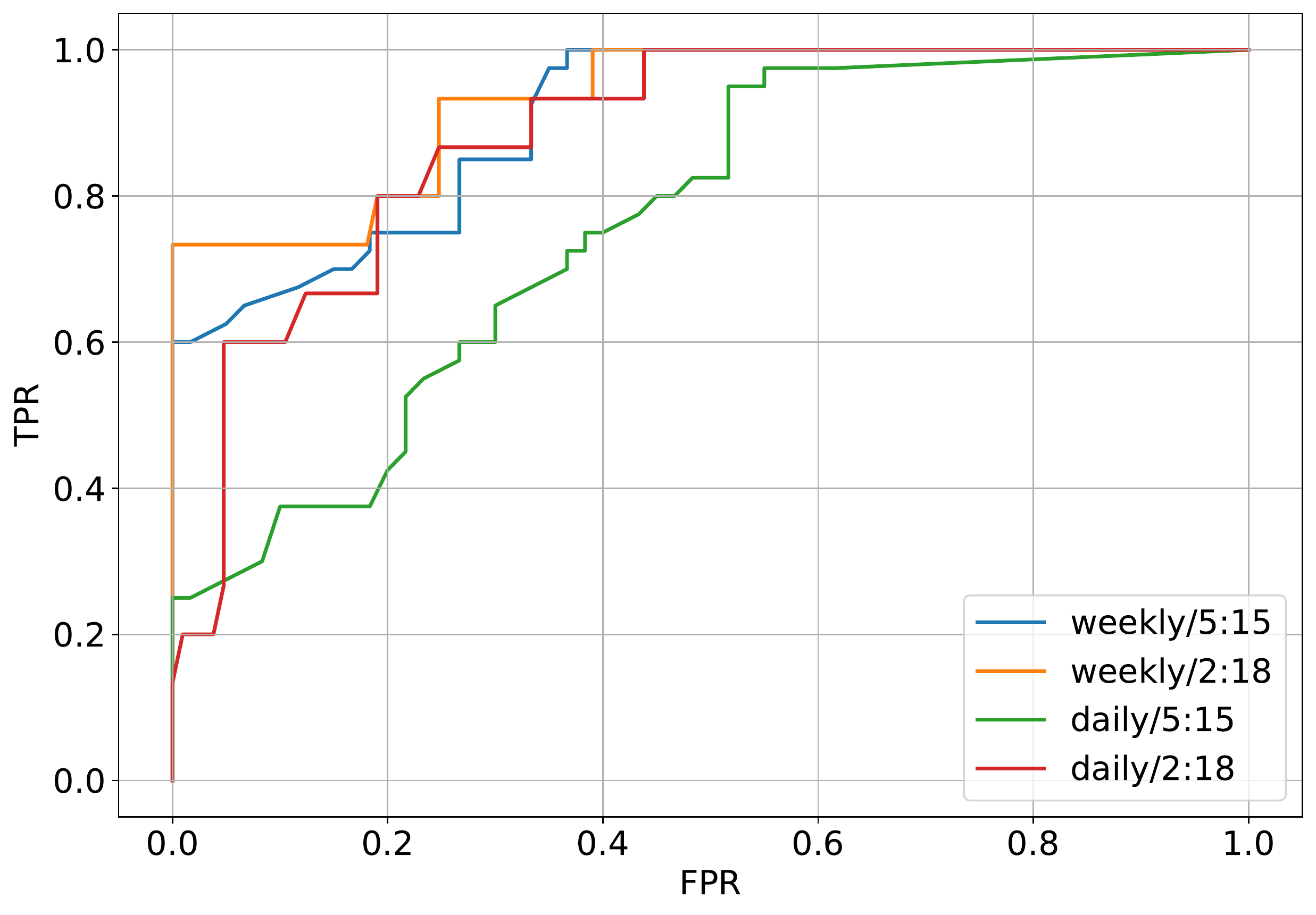}}	
	\subfigure[ROC curve]{
		\includegraphics[width=0.45\textwidth]{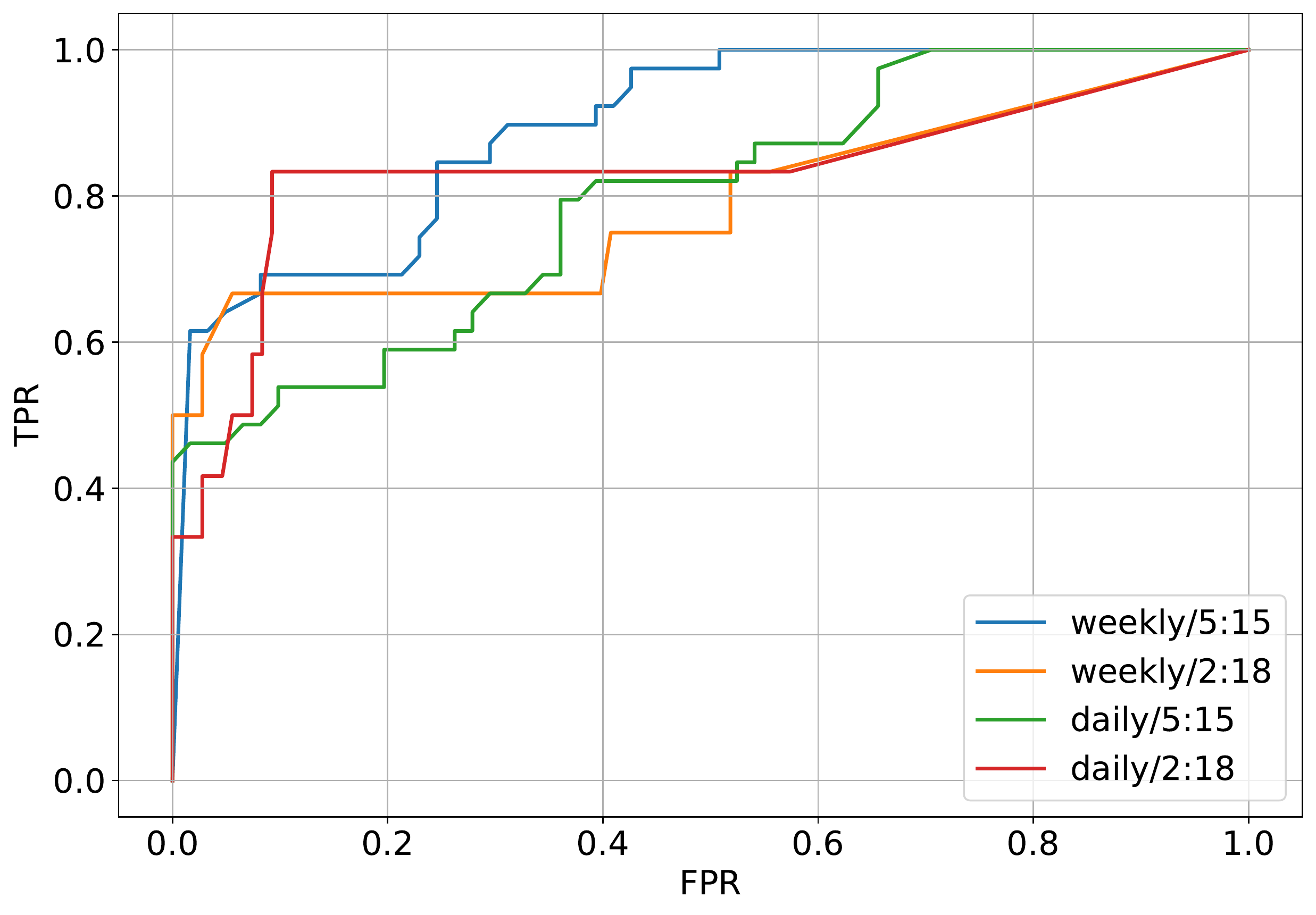}}	
		
	\caption{P-R and ROC curves for Task 2. Each plot is generated by randomly selecting one day out of the range April 1st, 2018---December 31st, 2019 and aggregating predictions for that day over all runs.}
	\label{fig:pr_appendix}
\end{figure*}

In this Appendix, we first show how we map the data from Twitter's Election Integrity archives~\cite{twitterElectionHub} to the source/target disinformation campaign labels used in the study (Table~\ref{tbl:dataset_mapping_appendix}). Next, in Figure~\ref{fig:appendix_activity}, we show the high-level coordinated activity of the remaining disinformation campaigns and baselines that are not shown in Figure~\ref{fig:activity}. Finally, Figure~\ref{fig:pr_appendix} shows the P-R curves.

\begin{table}
	\centering
	\begin{tabular}{|l|c|}
	\hline

	Disinformation Campaign & Release Date\\
	\toprule

    Unknown $\rightarrow$ Serbia                                     & April 2020\\
	\hline
	Turkey $\rightarrow$ Turkey                                      & June 2020\\
	\hline
	Saudi Arabia (external) $\rightarrow$ Saudi Arabia               &  April 2020\\
	\hline
	Saudi Arabia $\rightarrow$ Geopolitical                                    &  December 2019\\
	\hline
	Egypt $\rightarrow$ Iran/Quatar/Turkey                                           &  April 2020\\
	\hline
	Russia (internal) $\rightarrow$ Russia                                &  June 2020\\
	\hline
	Indonesia $\rightarrow$ Indonesia                                    &  February 2020\\
	\hline
    Honduras $\rightarrow$ Honduras                  &  April 2020\\
	\hline
	China (2020) $\rightarrow$ Hong Kong                               &  June 2020\\
	\hline
	Russia (external) $\rightarrow$ Russia                                    &  March 2020\\
	\bottomrule

	\end{tabular}
   \vspace{0.1in}
    \caption{This table shows how we mapped the data release from Twitter's Election Integrity Hub to the disinformation campaign in our study.}
 	\label{tbl:dataset_mapping_appendix}
\end{table}

\end{document}